\documentclass{jfm}
\usepackage{graphicx}
\usepackage{epstopdf, epsfig}
\usepackage{epstopdf, epsfig}
\usepackage{amsmath}
\usepackage{color}

\newcommand\calD{\mathcal{D}} 
\newcommand\Retau{{\textit{Re}_\tau}}  

\def\reffig#1{Fig.~\ref{fig:#1}}

\def\refsec#1{\S\ref{sec:#1}}
\def\refeq#1{Eq.~(\ref{eq:#1})}

\def\calD    {\mathcal{D}}

\def\state {\mathbf{q}}
\def\vel   {\mathbf{u}}
\def\forc  {\mathbf{f}}

\def\statehat {\hat{\state}}
\def\velhat   {\hat{\vel}}
\def\forchat  {\hat{\forc}}

\def\veltil   {\widetilde{\vel}}
\def\forctil  {\widetilde{\forc}}

\def\Scorr    {\mathbf{S}}
\def\Scorruu  {{S_{uu}}}
\def\Pcorr    {\mathbf{P}}

\def\bfx   {\mathbf{x}}

\def\bfA   {\mathbf{A}}
\def\bfB   {\mathbf{B}}
\def\bfC   {\mathbf{C}}
\def\bfD   {\mathbf{D}}
\def\bfR   {\mathbf{R}}
\def\bfP   {\mathbf{P}}

\def\bfI   {\mathbf{I}}
\def\bfH   {\mathbf{H}}

\def\bfU   {\mathbf{U}}

\definecolor{mydarkgreen}{RGB}{0,90,20}

\def\avg#1{\langle{#1}\rangle}

\def\spe{power spectral density}
\def\crospe{{power cross-spectral density}}

\shorttitle{Relevance of Reynolds stresses in resolvent analysis}
\shortauthor{P. Morra, O. Semeraro, D.S. Henningson \& C. Cossu}

\title{On the relevance of Reynolds stresses\\ in resolvent analyses of turbulent\\ wall-bounded flows}

\author{Pierluigi Morra\aff{1},
 Onofrio Semeraro\aff{2},
 Dan S. Henningson\aff{1},\\
 \and Carlo Cossu\aff{3}
  \corresp{\email{carlo.cossu@ec-nantes.fr}}
}

\affiliation{ 
\aff{1} KTH Royal Institute of Technology, Linn\'e FLOW Centre, SE-10044, Stockholm, Sweden
\aff{2} LIMSI, UPR 3251 CNRS / Universit\'{e} Paris-Saclay, 91400 Orsay, France
\aff{3} LHEEA, UMR 6598  CNRS / Centrale Nantes, 44300 Nantes, France 
}

\begin{document}

\maketitle

\begin{abstract}
The ability of linear stochastic response analysis to estimate coherent motions is investigated in turbulent channel flow at friction Reynolds number $\Retau=1007$.
The analysis is performed for spatial scales characteristic of buffer-layer and large-scale motions by separating the contributions of different temporal frequencies.
Good agreement between the measured spatio-temporal power spectral densities and those estimated by means of the resolvent 
is found  when the effect of turbulent Reynolds stresses, modelled with an eddy-viscosity associated to the turbulent mean flow, is included in the resolvent operator. 
The agreement is further improved when the flat forcing power spectrum (white noise) is replaced with a power spectrum matching the measures.
Such a good agreement is not observed when the eddy-viscosity terms are not included in the resolvent operator. 
In this case, the estimation based on the resolvent is unable to select the right peak frequency and wall-normal location of buffer-layer motions.
Similar results are found when comparing truncated expansions of measured streamwise velocity power spectral densities  based on a spectral proper orthogonal decomposition to those obtained with optimal resolvent modes.
\end{abstract}


\section{Introduction}
\label{sec:intro}

Most of the fluctuating energy in wall-bounded turbulent shear flows resides in highly correlated streamwise velocity fluctuations associated with coherent streamwise streaks, i.e. spanwise alternated high- and low-velocity regions elongated in the streamwise direction. 
The existence of these structures is known since the visualisations of \cite{Kline1967} which revealed that the near-wall region of turbulent boundary layers is populated by streaks with average spanwise spacing  $\lambda_z^+ \approx 100$ in the buffer layer and higher farther from the wall.
Streaky motions with larger scales have been observed in the logarithmic and the outer region where `large-scale' (LSM) and `very large scale' motions (VLSM) have typical spanwise spacings $\lambda_z \approx \delta - 1.5 \delta$ (where $\delta$ is the outer length scale) and streamwise size of $\lambda_x \approx 2\lambda_z$ \citep{Corrsin1954,Kovasznay1970} and $\lambda_x \simeq O(10\lambda_z)$  \citep{Komminaho1996,Kim1999,Hutchins2007}, respectively.

The ubiquity of streaks in transitional and turbulent flows has been related to the `lift-up' effect where high-energy streamwise streaks are induced by low-energy quasi-streamwise vortices immersed in a shear flow \citep{Moffatt1967,Ellingsen1975,Landahl1980} via algebraic growth \citep{Ellingsen1975,Gustavsson1991}. 
The large energy amplifications associated with this mechanism have been related to the non-normality of the linearised Navier-Stokes operator \citep[e.g][]{Boberg1988,Reddy1993,Trefethen1993} and much attention has been given to the computation of the largest (optimal) energy amplifications supported by laminar basic flows \citep{Gustavsson1991,Butler1992,Reddy1993,Schmid2001}.
\cite{Boberg1988} suggested that the subcritical onset of turbulence in wall-bounded flows can be attributed to a mechanism where the low-energy quasi-streamwise vortices 
leading to the amplification of high-energy streamwise streaks are then regenerated by nonlinear effects related to the breakdown of the  streaks.
A similar `self-sustained mechanism' has been invoked by \cite{Hamilton1995} to explain the dynamics of buffer-layer streaks in the turbulent regime with \cite{Waleffe1995} and \cite{Reddy1998} attributing the breakdown of the streaks to a modal secondary instability and \cite{Schoppa2002} to a secondary non-modal energy amplification. 
\cite{Hwang2010b,Hwang2011} have found evidence that similar {\it coherent} self-sustained processes sustain all active streaky motions, with scales ranging from those of buffer-layer streaks to those of large-scale motions. 

While a relative consensus begins to emerge about the existence of self-sustained processes, there are different interpretations of the mechanisms involved in these processes in turbulent flows, and, in particular, of the nature of the linear operator intervening in the non-normal amplification mechanism. 
In a first approach, pursued by \cite{Malkus1956,Butler1993,Farrell1993,McKeon2010} and many others, the Navier-Stokes equations are rewritten in terms of perturbations to the turbulent mean velocity; the instantaneous and averaged (Reynolds stresses) perturbation nonlinear terms are accounted for as an external input which forces the response via the linear operator.
In the second approach, initiated by \cite{Reynolds1967,Reynolds1972} and pursued by \cite{Bottaro2006,delAlamo2006,Cossu2009,Pujals2009,Hwang2010,Hwang2010c} among others, the `incoherent' part of turbulent Reynolds stresses is included in the linear operator by means of a $\nu_t$  eddy-viscosity modelling. 
We will refer to the latter approach as `$\nu_t$-model' and to the former as `$\nu$-model'.

An important number of investigations have dealt with the computation of optimal energy amplifications and of the associated optimal inputs and outputs for the linear initial-value problem and the response to harmonic and stochastic forcing.
While a review of this research effort and of the respective merits attributed to the $\nu$ and $\nu_t$ formulations is beyond the scope of this paper \citep[we refer the reader to][for a summary of recent results]{McKeon2017,Cossu2017}, it appears that most of the comparisons between linear models predictions and real turbulent flows are either qualitative or concern wall-normal integrated energy densities and the reproduction of spatial spectra (the Fourier transform of the second-order velocity spatial correlations). Also, with very few exceptions \citep[e.g.][]{Illingworth2018}, these analyses very often focus only on the most amplified modes without resulting in a detailed quantitative comparison of the performance of the $\nu$ and $\nu_t$ models.

A similar situation has emerged in the evaluation of the performance of Karhunen-Lo\`eve decompositions (proper orthogonal decompositions, POD) in the approximation of the \spe.
In this context it has been recognised that the aggregation of all frequencies in the analysis, typically used when only spatial correlations are estimated, often blurs the interpretation of the results.
It has hence been suggested to analyse separately each contributing frequency by making use of the Fourier transform of the {\it spatio-temporal} correlation tensor $\bfR(\bfx,\bfx+\Delta \bfx,t,t+\tau)$ and the associated `spectral' POD modes \citep{Lumley1970,Picard2000}.
It has then been highlighted that a strong relation exists between these spectral POD (SPOD) modes and those issued from the Schmidt (singular value, SVD) decomposition of the resolvent operator \citep{Semeraro2016b,Towne2018}. 
However, in those investigations, only the performance of the resolvent analysis based on the $\nu$-model was evaluated.

The scope of the present study, inspired by the approach recently revived in the context of POD analyses, is therefore to evaluate the respective performance of the $\nu$-model and the $\nu_t$-model in the estimation of the velocity \emph{spatio-temporal} \spe\, and \crospe.
The turbulent channel flow at $\Retau=1007$ is used as a testbed for this analysis.

The paper is organized as follows: the mathematical formulation of the problem addressed is briefly summarized in \refsec{bground}.
The turbulent channel flow obtained by direct numerical simulation and the relevant spectral information are described in \refsec{DNS} and compared to the estimations obtained via the resolvent-based models in \refsec{LinMods}.
The results are summarized and discussed in \refsec{concl}.
Further details on the operators involved in the resolvent-based models and on the data analysis of direct numerical simulations data are provided in App.~A and B, respectively.

\section{Background} \label{sec:bground}

\subsection{Linear models for the dynamics of turbulent fluctuations}

We are interested in the dynamics of coherent perturbations in a pressure-driven turbulent channel flow of incompressible viscous fluid with density $\rho$, kinematic viscosity $\nu$ between two infinite parallel walls located at $y=\pm h$. Here, we denote by $x$, $y$ and $z$ the streamwise, wall-normal and spanwise coordinates, respectively. 
In the following we will consider dimensionless variables based on the reference length $h$ and the reference velocity $(3/2) U_{bulk}$, where $U_{bulk}$ is the constant mass-averaged streamwise velocity.

Following \cite{Reynolds1967,Reynolds1972,delAlamo2006,Pujals2009} and many others, 
the evolution of small coherent perturbations to the turbulent mean flow can be modelled with linearised equations which include the effect of the turbulent Reynolds stresses in the linear model by means of an eddy viscosity $\nu_t$ corresponding to the turbulent mean flow profile $\bfU=(U(y),0,0)$:
\begin{equation}
\label{eq:LinNS}
\frac{\partial \vel}{\partial t} + \nabla \vel \cdot \bfU+\nabla \bfU \cdot \vel
 =  -{1 \over \rho} \nabla p + \nabla \cdot \left[ \nu_T \left(\nabla \vel +\nabla \vel^T \right) \right]+\mathbf{f},
\end{equation}
where $\vel=(u,v,w)$ and $p$ are the coherent perturbation velocity and pressure, $\nu_T(y)=\nu+\nu_t(y)$ is the total effective viscosity and  $\mathbf{f}$ is the forcing term. 
These equations are completed by the  incompressibility condition $\nabla\cdot\vel=0$.

For the eddy viscosity, we adopt the semi-empirical expression proposed by \cite{Cess1958}, as reported by \cite{Reynolds1967}:
\begin{equation}
\nu_t = {\nu \over 2} \big\{1+{{\kappa^2 \Retau^2} \over 9}(1-y^2)^2(1+2 y^2)^2 
(1-e^{y^+/A})^2\big\}^{1/2}-{\nu\over 2},
\end{equation}
where  $y\in[-1,1]$ is the centre-channel based dimensionless wall-normal coordinate, $y^+=\Retau(1-|y|)$ and $\Retau=u_\tau h /\nu$ is the Reynolds number based on the friction velocity $u_\tau$. We set the von K\'{a}rm\'{a}n constant $\kappa=0.426$ and the constant $A=25.4$ as in \cite{Pujals2009,Hwang2010c}. 
In the following we will refer to this model as the `$\nu_t$-model'.

In an alternative approach, used e.g. by \cite{Butler1993,Farrell1993b,McKeon2010} and many others, turbulent Reynolds stresses and nonlinear terms are both included in the forcing term $\forc$ (which is therefore different from the forcing term of the $\nu_t$-model) so that the linear model reduces to \refeq{LinNS} but now only including the molecular kinematic viscosity $\nu_T=\nu$. 
We will refer to this model as `$\nu$-model' (also known as `quasi-laminar model').

\subsection{Analyses of harmonic and stochastic forcing}

Since the mean flow is homogeneous in the streamwise and spanwise directions,  we consider the Fourier-modes ${\hat{\vel}}(\alpha,y,\beta,t) e^{i(\alpha x +\beta z)}$ and ${\hat{\forc}}(\alpha,y,\beta,t) e^{i(\alpha x +\beta z)}$ of streamwise and spanwise wavenumbers $\alpha=2 \pi/\lambda_x$ and $\beta=2 \pi/\lambda_z$.
\refeq{LinNS} can be reduced to the following linear system, expressed in terms of the state vector $\statehat = \left[\hat{v} , \hat{\omega}_y \right]^T$ formed with the wall-normal components of the velocity and vorticity Fourier modes:
\begin{eqnarray}
{\partial \statehat \over \partial t} = \bfA \statehat + \bfB \forchat,
\label{eq:DynSys}
\end{eqnarray}
where $\velhat = \bfC \statehat$ and $\statehat = \bfD \velhat$ (the explicit expressions of the operators $\bfA$, $\bfB$, $\bfC$ and $\bfD$ are given in App.~A).
As the system~(\ref{eq:DynSys}) is linearly stable \citep{Reynolds1967}, the response to deterministic finite-power forcing can be analysed by Fourier-transforming the linear system in time, hence considering the harmonic forcing $\forchat=\forctil e^{-i \omega t}$ with the harmonic response  $\veltil e^{-i \omega t}$ related by:
\begin{eqnarray}
\label{eq:H}
\veltil &=& \bfH \forctil~~~~;~~~~ \bfH=-\bfC (i \omega \bfI + \bfA)^{-1} \bfB
\label{eq:T}
\end{eqnarray}
where $\bfH$ is the resolvent operator (or transfer function).
When system ~(\ref{eq:DynSys}) is driven by stochastic forcing, the response is also stochastic.
For the considered horizontally- and time-invariant problem the velocity second-order {\it spatio-temporal} correlation tensor is defined as
\begin{equation}
\bfR(\xi, y, y',\zeta,\tau) = \langle{{\mathbf{u}}(x,y,z,t) {\mathbf{u}}^*(x+\xi,y',z+\zeta,t+\tau)}\rangle,
\end{equation} 
where $\avg{~}$ denotes ensemble averaging and $^*$ denotes complex conjugate transpose. 
The (spatio-temporal) \crospe\, tensor is obtained through Fourier transform in the horizontal plane and time: 
\begin{eqnarray}
\Scorr(\alpha,y,y',\beta,\omega)  = \frac{1}{(2 \pi)^3} \int_{-\infty}^\infty \int_{-\infty}^\infty \int_{-\infty}^\infty \bfR(\xi, y, y',\zeta,\tau)\, e^{i(\alpha \xi + \beta \zeta - \omega \tau)} d\xi d\zeta d\tau.
\end{eqnarray}
The velocity \crospe\, tensor $\Scorr$ can also be obtained directly as the average of the Fourier transform $\veltil$ of the velocity \citep[e.g.][]{Bendat1986} and the stochastic forcing \crospe\, tensor $\Pcorr$ can be defined similarly:
\begin{eqnarray}
\Scorr = \avg{\veltil(\alpha,y,\beta,\omega) \veltil^*(\alpha,y',\beta,\omega)}
~~~~;~~~~ 
\Pcorr = \avg{\forctil(\alpha,y,\beta,\omega) \forctil^*(\alpha,y',\beta,\omega)}.
\label{eq:S}
\end{eqnarray}
An estimation $\Scorr^{(est)} = \bfH \Pcorr \bfH^*$ of the velocity \crospe\, tensor is obtained by making use of the linear models by replacing the linear expression of \refeq{H} in \refeq{S}.
If it is assumed that $\forctil$ is $\delta$-correlated in the wall-normal direction $y$, with decorrelated components 
\citep[see e.g.][]{Farrell1993,Hwang2010,Hwang2010c}, then  
$\Pcorr=p(\alpha,\beta,\omega)\, \bfI$ (where $\bfI$ is the identity operator).
This assumption, which has been made in a number of previous studies, underlies the so-called resolvent analysis because in this case  $\Scorr^{(est)}$ reduces to:
\begin{eqnarray}
 \Scorr^{(res)} = p \bfH \bfH^*
 \label{eq:Sres}
\end{eqnarray}
The scope of this study is to explore the validity of the estimation provided by \refeq{Sres}.

\begin{figure}
  \centering
   \centerline{
  \includegraphics[width=0.35\textwidth]{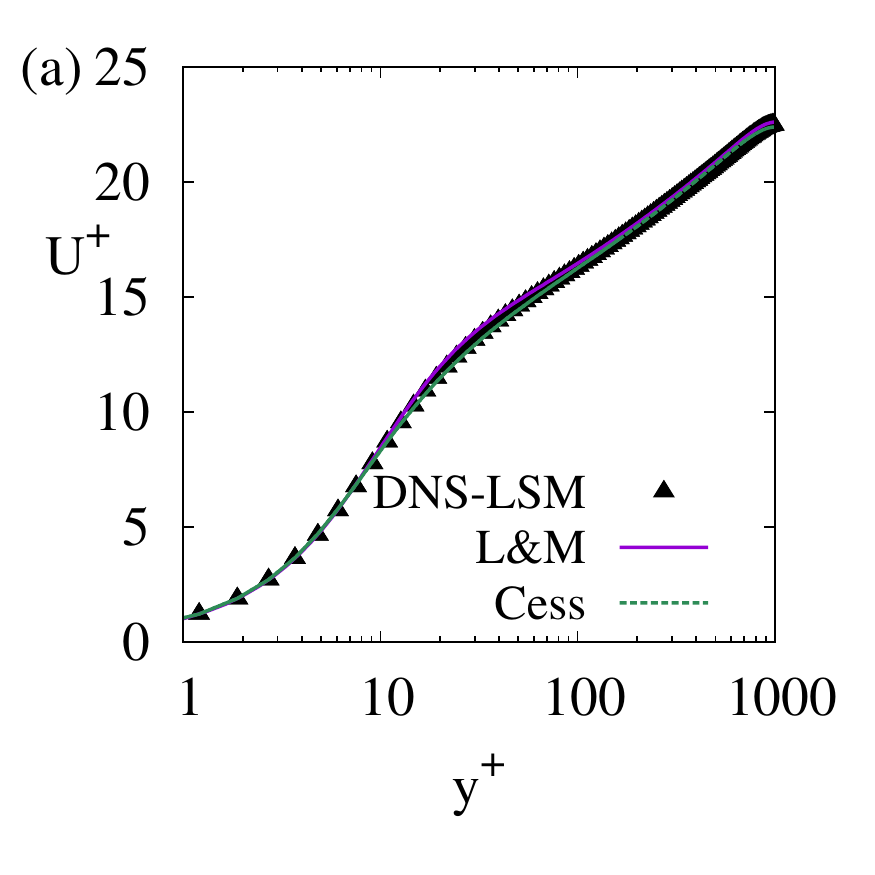}\hspace{-4mm}
  \includegraphics[width=0.35\textwidth]{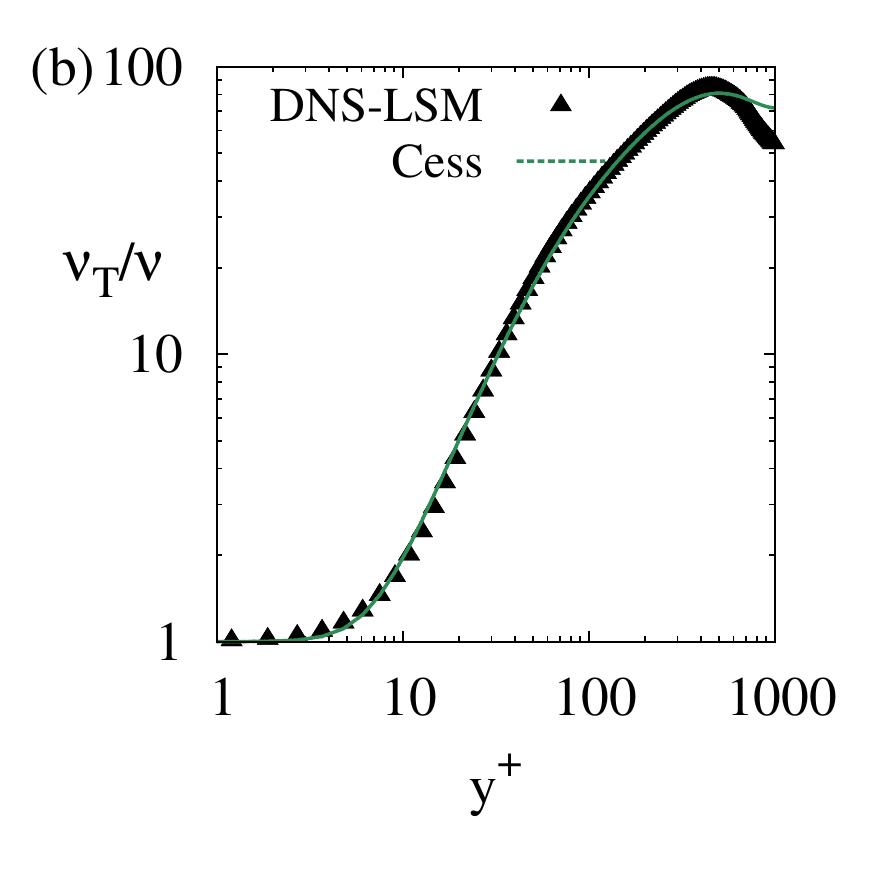}\hspace{-4mm}
  \includegraphics[width=0.35\textwidth]{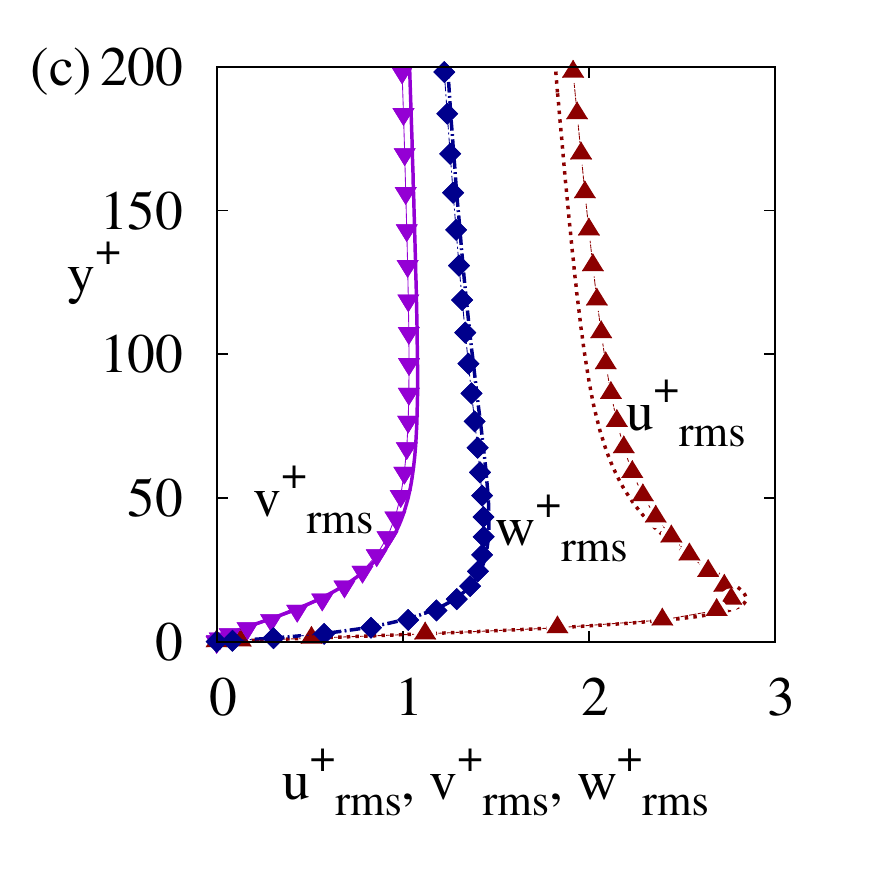}
}
  \vspace*{5mm}
  \caption{Comparison of direct numerical simulations in the LSM flow unit (DNS-LSM) at $\Retau=1007$  to those (L\&M) of \cite{Lee2015} and to Cess's model at $\Retau=1000$: 
$(a)$~mean flow profile, 
$(b)$~effective turbulent viscosity $\nu_T=\nu_t+\nu$ corresponding to the mean flow profile and 
$(c)$~$rms$ velocity profiles from the DNS-LSM (lines with symbols) compared to those of L\&M (lines).
} \label{fig:MeanFlow}
\end{figure}
\begin{figure}
  \centering
   \centerline{
  \includegraphics[width=0.39\textwidth]{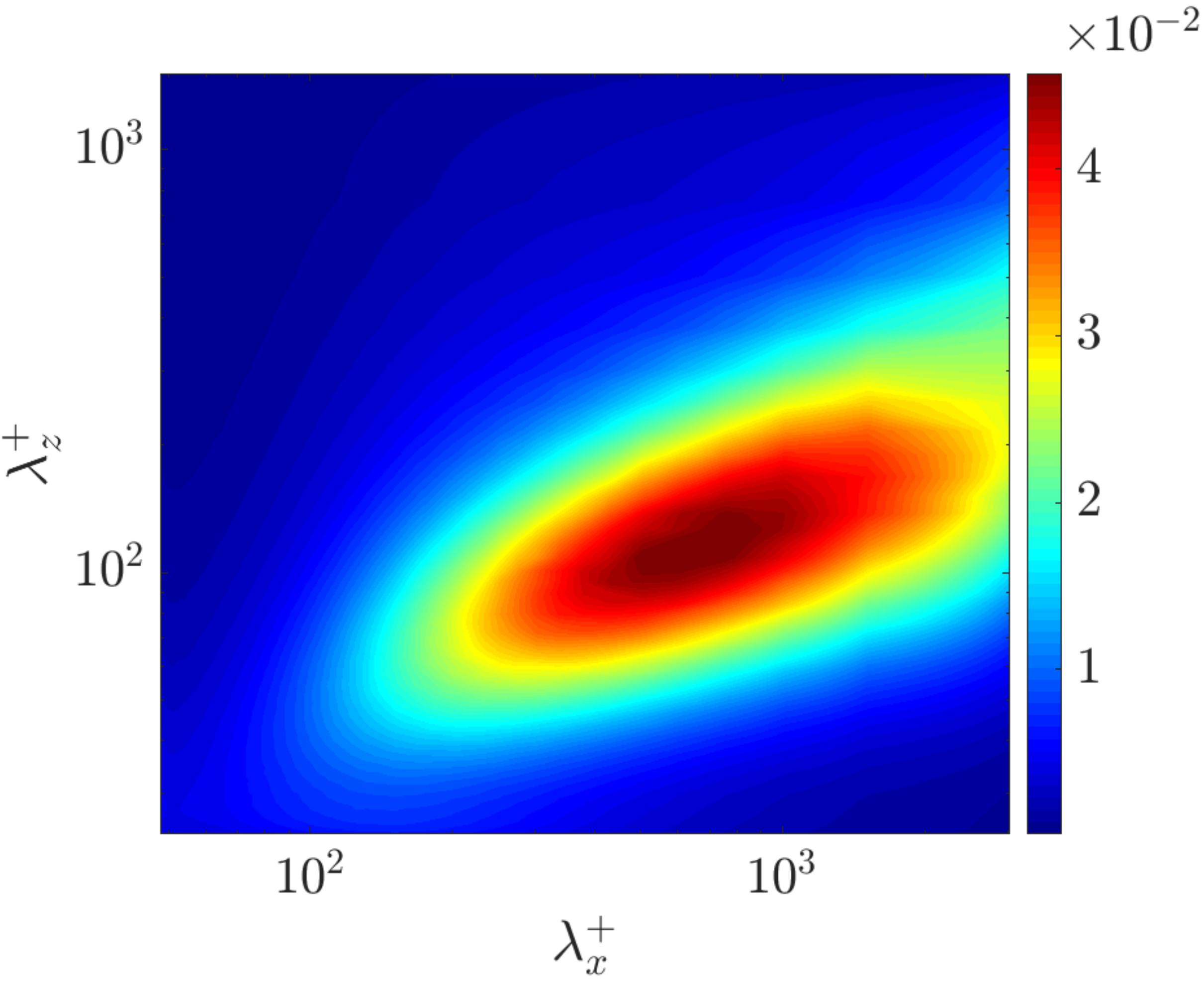}~~~~~~~
\put(-170,120){{$(a)$}} 
  \includegraphics[width=0.39\textwidth]{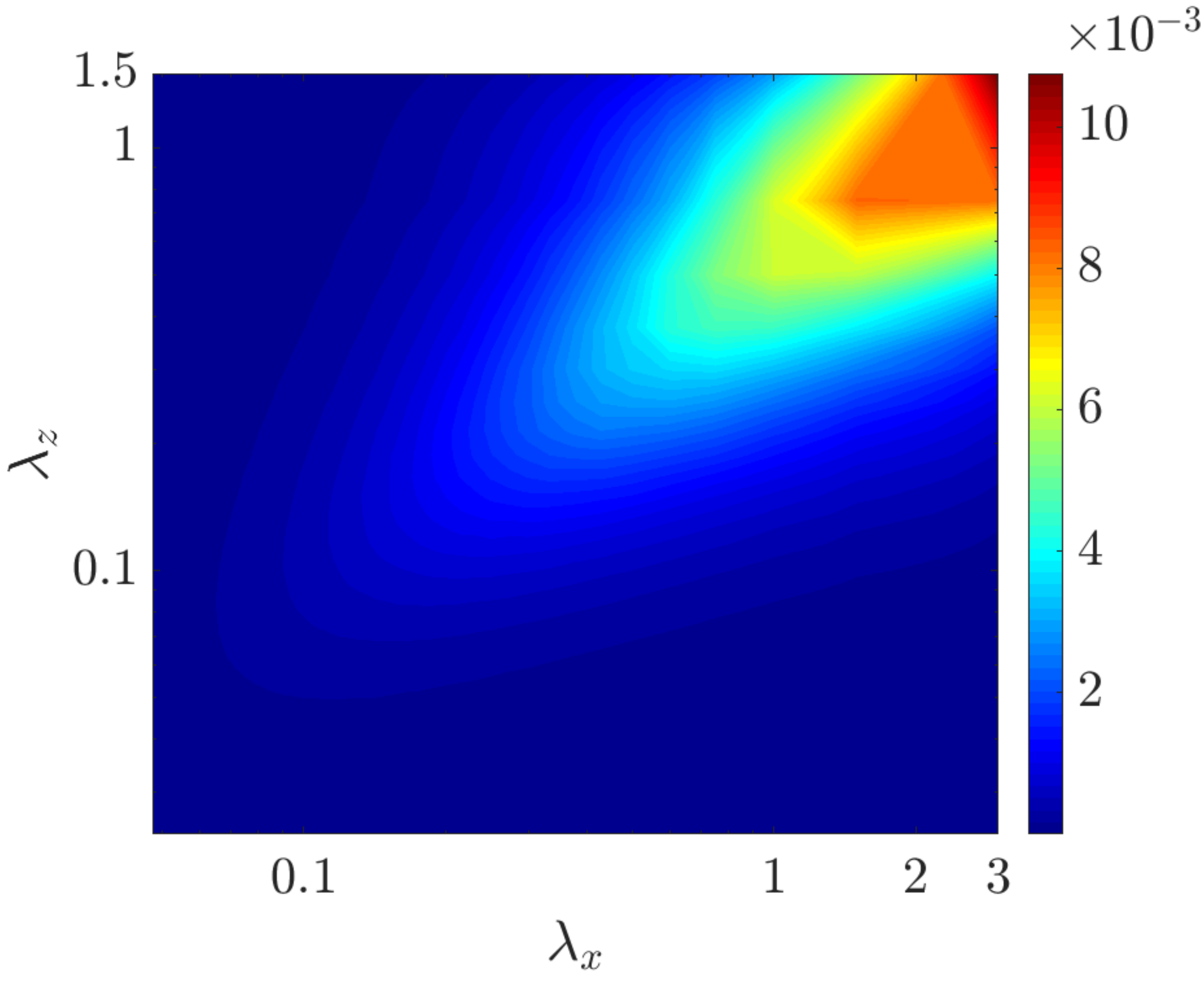} 
\put(-155,120){{$(b)$}}
}  
  \caption{Premultiplied streamwise energy spectra $\alpha \beta E_{uu}$ in the $y^+=15$ plane (panel $a$) and the $y=0.5$ plane  (panel $b$) . Data from the DNS in the LSM flow unit at $\Retau=1007$.
} \label{fig:PreSpe}
\end{figure}

\section{Direct numerical simulations in a LSM flow unit}
\label{sec:DNS}

\begin{figure}
  \centering
   \centerline{
 \includegraphics[width=0.39\textwidth]{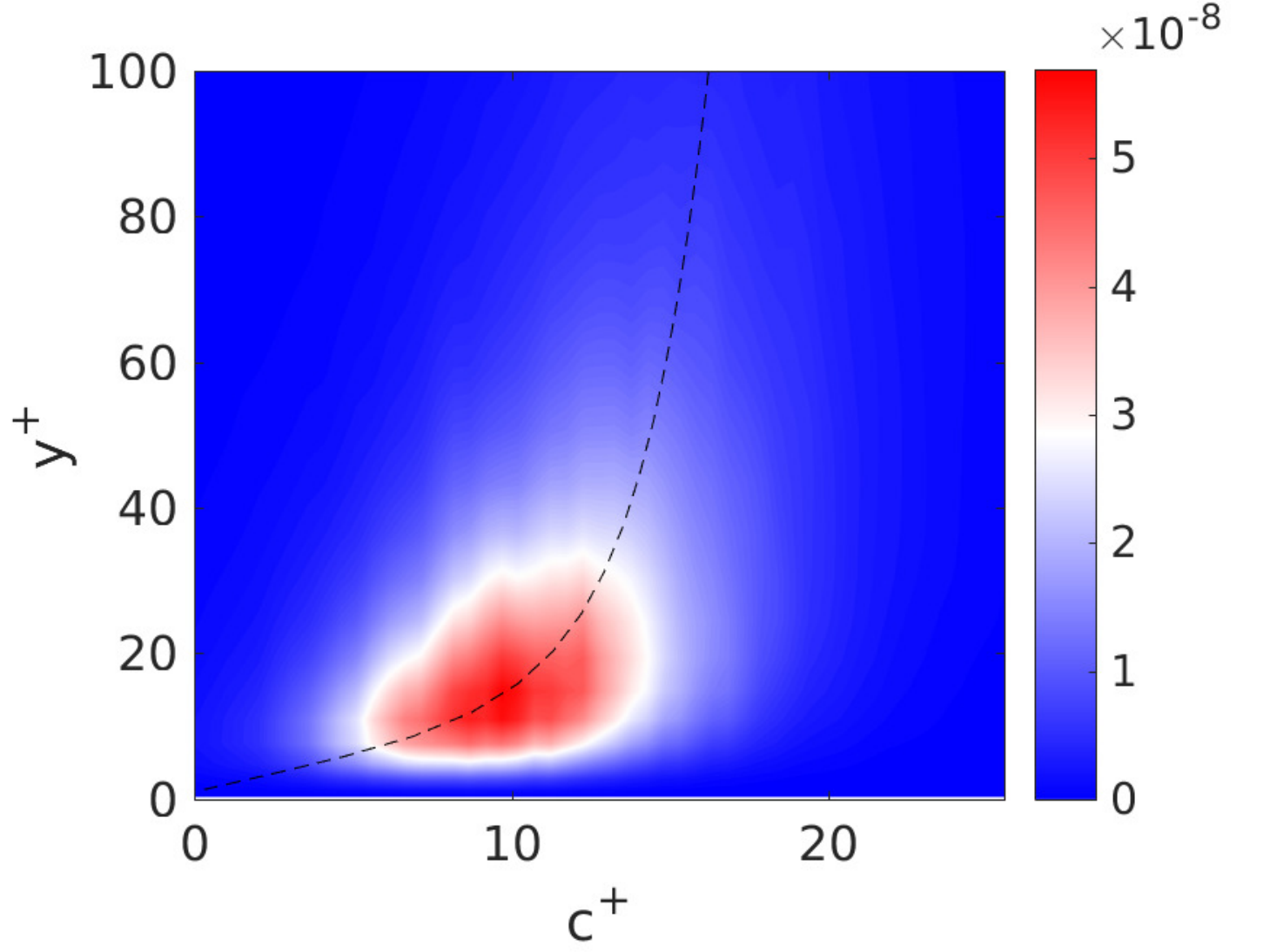} 
\put(-155,100){{$(a)$}}
 \includegraphics[width=0.39\textwidth]{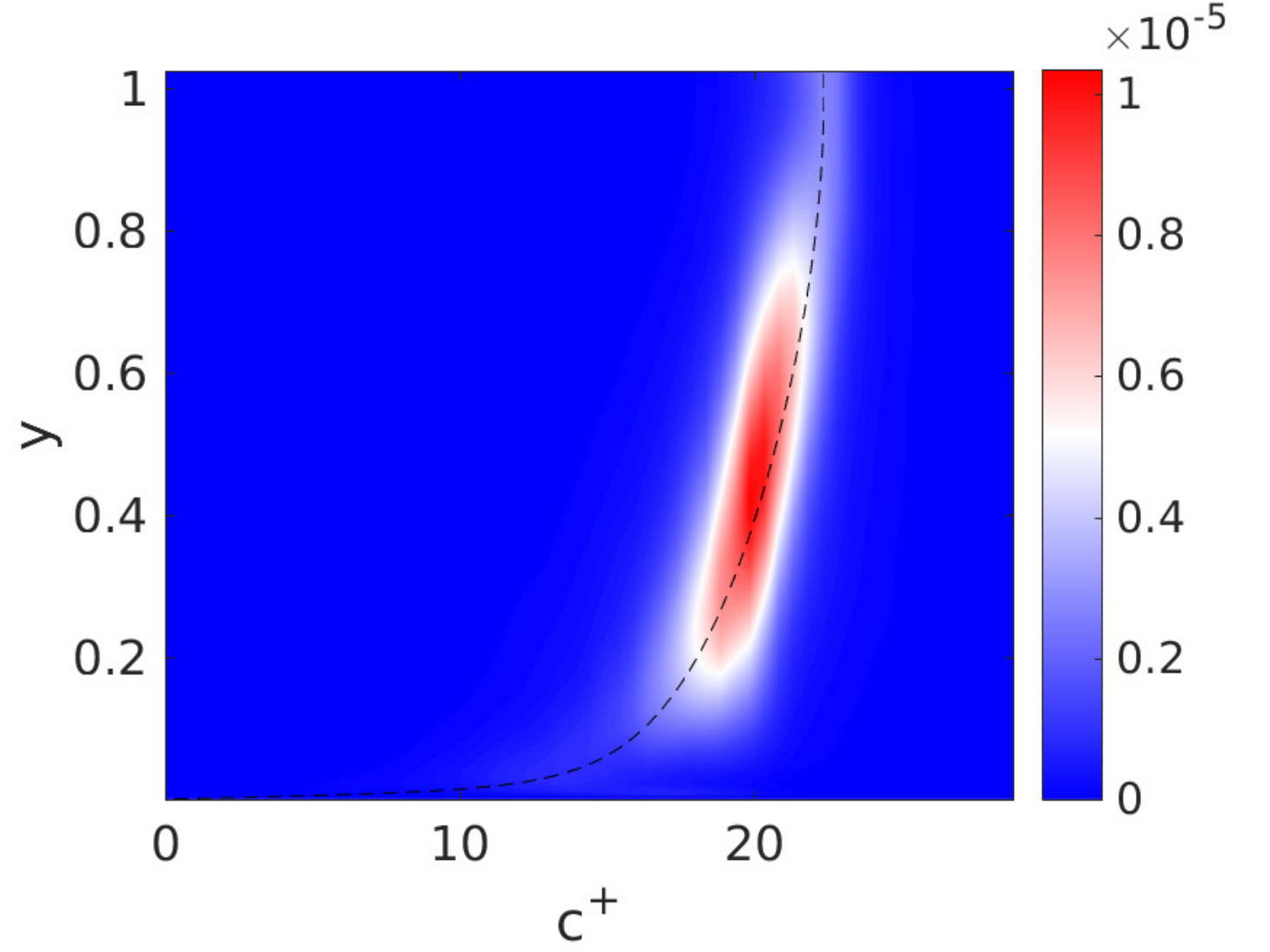}
\put(-150,100){{$(b)$}} 
}  
\caption{Wall-normal profiles of the streamwise velocity \spe\, 
$\avg{\widetilde{u}(\alpha,y,\beta,\omega) \widetilde{u}^*(\alpha,y,\beta,\omega)}$ versus the phase speed $c^+=\omega^+/\alpha^+$ for 
$(a)$~buffer-layer structures with $\lambda_x^+ = 450,\lambda_z^+ = 100$ and 
$(b)$~large-scale structures with $\lambda_x=3, \lambda_z=1.5$.
The mean velocity profile $U^+$ is reported as a dashed black line.
Data from the direct numerical simulation in the LSM flow unit at $\Retau=1007$.}
\label{fig:Welch}
\end{figure}
\begin{figure}
  \centerline{
  \includegraphics[width=0.39\textwidth]{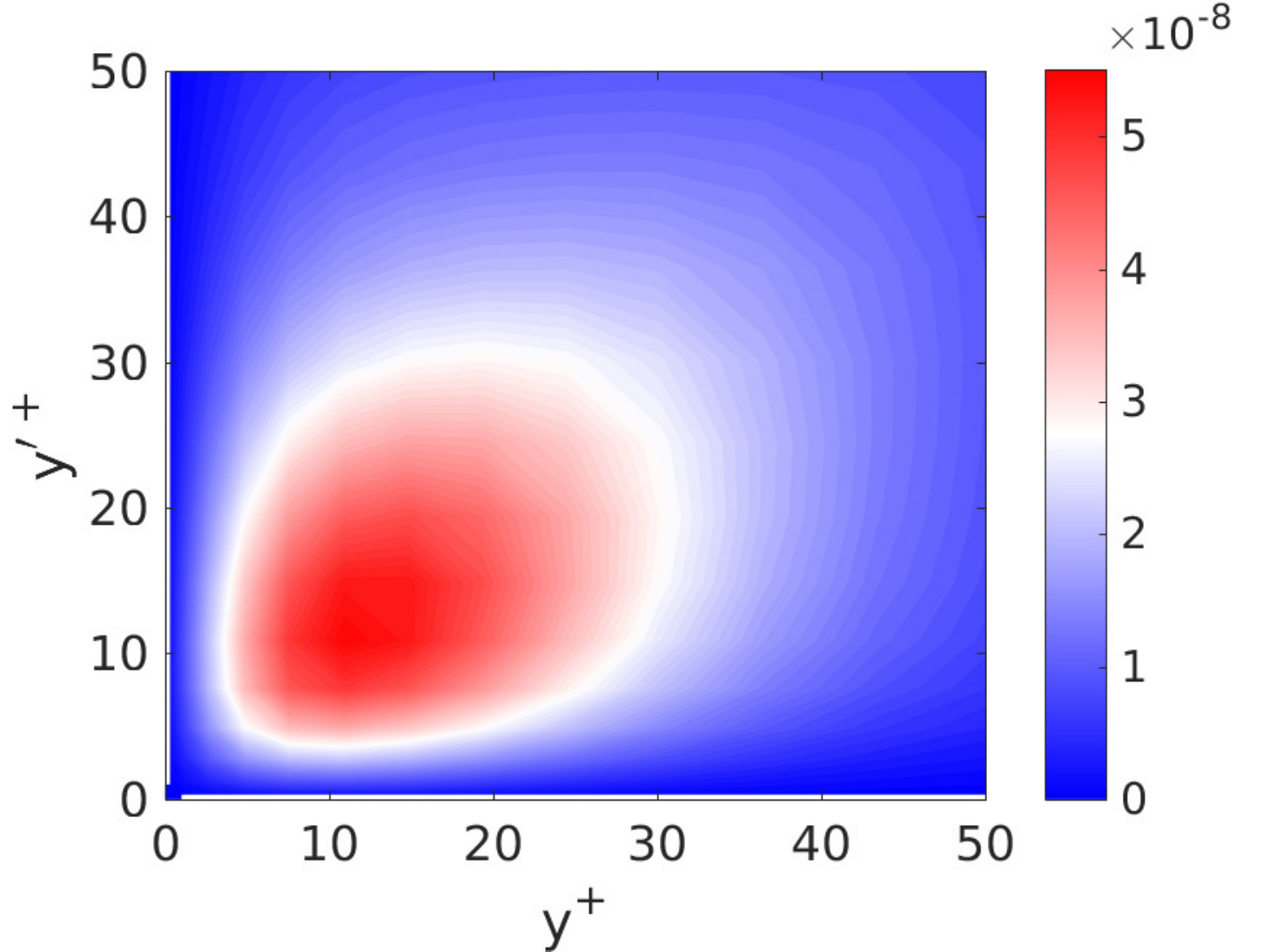} 
\put(-155,100){{$(a)$}}
  \includegraphics[width=0.39\textwidth]{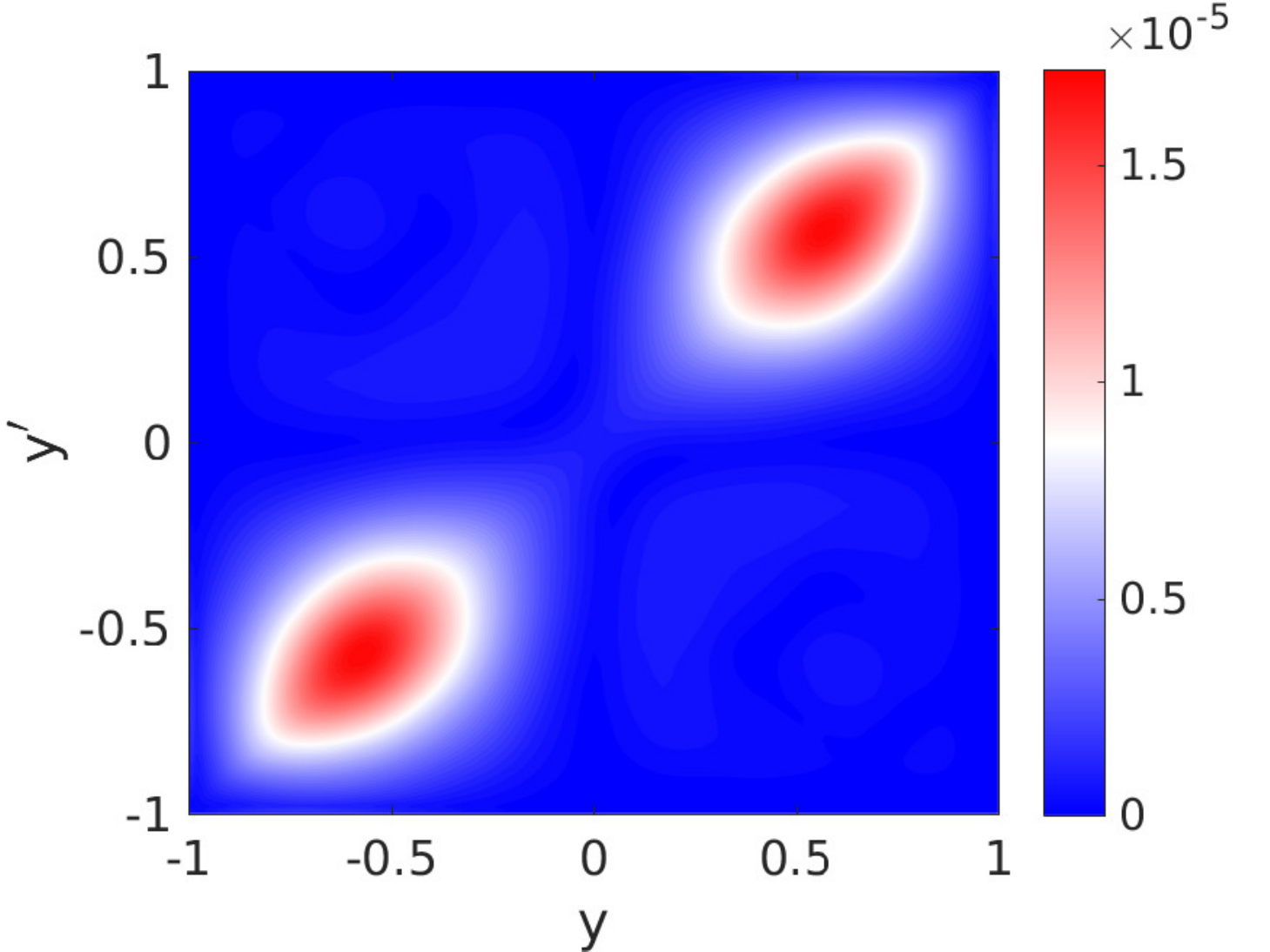}
\put(-150,100){{$(b)$}}
}
\caption{Absolute value of the streamwise velocity \crospe\,
$|\avg{\widetilde{u}(\alpha,y,\beta,\omega_{max}) \widetilde{u}^*(\alpha,y',\beta,\omega_{max})}|=
|\Scorruu(\alpha,y,y',\beta,\omega_{max})|$ for 
$(a)$~near-wall structures with 
$\lambda_x^+ = 450,\lambda_z^+ = 100$ at the \spe\, peak value 
$\omega_{max}=4.3$ and 
$(b)$~large-scale structures with $\lambda_x=3, \lambda_z=1.5$ at the \spe\, peak value $\omega_{max}=1.4$.
Data from the direct numerical simulation in the LSM flow unit at $\Retau=1007$.
} 
\label{fig:WelchCrossCo}
\end{figure}

The turbulent channel flow has been simulated at $\Retau=1007$ in a domain $L_x=3$ long and  $L_z=1.5$ wide 
in order to focus on the physics of the near-wall and large-scale processes only and to keep the data processing manageable.
This domain size also corresponds to that of the most energetic large-scale motions (LSM) in the channel \citep{delAlamo2004} and represents the minimum flow unit for the self-sustainment of coherent LSM \citep{Hwang2010b}.

We have verified that the computed mean flow and the $rms$ fluctuation profiles are in reasonable agreement with those of \cite{Lee2015} obtained in the much larger domain $L_x= 25, L_z= 9.5$ as shown in \reffig{MeanFlow}.
We have further verified that Cess's analytic fit works well by comparing the mean flow profile and the corresponding eddy viscosity to the ones issued by DNS data.
It should be remarked that the values of the eddy viscosity are small in the near-wall region but very large in the bulk of the flow where $\nu_T/\nu \sim O(10-100)$ at the considered $\Retau$.

The premultiplied streamwise kinetic energy spectral density $\alpha \beta E_{uu}(\alpha,y,\beta)=\alpha \beta \int_{-\infty}^\infty \Scorruu(\alpha,y,y,\beta,\omega) d\omega$, evaluated in the $y^+=15$ and $y=0.5$ planes, is shown in \reffig{PreSpe}.
As expected, the most energetic structures at $y=0.5$ have spatial scales corresponding to the dimensions of the LSM flow unit  $\lambda_x = L_x = 3$, $\lambda_z = L_z = 1.5$  (corresponding to $\alpha \approx 2$, $\beta \approx 4$). 
The most energetic structures at $y^+=15$ have spatial scales $\lambda_x^+ \approx 450$ and $\lambda_z^+  \approx 100$ (corresponding to $\alpha \approx 14$, $\beta \approx 63$ at $\Retau=1007$) typical of the near-wall self-sustained process. 
In the following we will therefore concentrate on these two sets of wavenumbers $(\alpha,\beta)$, corresponding to $(\lambda_x=3, \lambda_z=1.5)$ and $(\lambda_x^+ = 450,\lambda_z^+ = 100)$, which represent the dynamics of large-scale and near-wall self-sustained motions, respectively.

The \crospe\, tensor has been computed for the two considered $(\alpha,\beta)$ pairs by means of Welch's method \citep[see e.g.][]{Bendat1986}, as detailed in App.~\ref{app:DNSmeth}. 
In \reffig{Welch} we report the streamwise velocity \spe\, profiles
$\Scorruu(\alpha,y,y,\beta,\omega)=\avg{\widetilde{u}(\alpha,y,\beta,\omega) \widetilde{u}^*(\alpha,y,\beta,\omega)}$ as a function of the phase speed $c=\omega/\alpha$, expressed in wall units, and of the wall normal coordinate for the considered $(\alpha,\beta)$ pairs.
The peaks of these distributions
are found at $\omega_{max}=4.3$ and $\omega_{max}=1.4$ for the near-wall and large-scale peaks, respectively. 
These peaks correspond to the two phase velocities 
$c^+_{max} \approx 10$ and $c^+_{max} \approx 20$ and their wall-normal locations ($y^+_{max}\approx 15$ and $y_{max}\approx 0.4-0.5$) approximatively correspond to the wall-normal position where $U^+=c^+$.

The $\Scorruu(\alpha,y,y',\beta,\omega)=\avg{\widetilde{u}(\alpha,y,\beta,\omega) \widetilde{u}^*(\alpha,y',\beta,\omega)}$ streamwise velocity power cross-spectral densities,
 are reported, for later comparison, in \reffig{WelchCrossCo} as a function of $y$ and $y'$ for the considered $(\alpha,\beta)$ at the peak $\omega_{max}$ corresponding to each pair.

\begin{figure}
  \centerline{
 \includegraphics[width=0.39\textwidth]{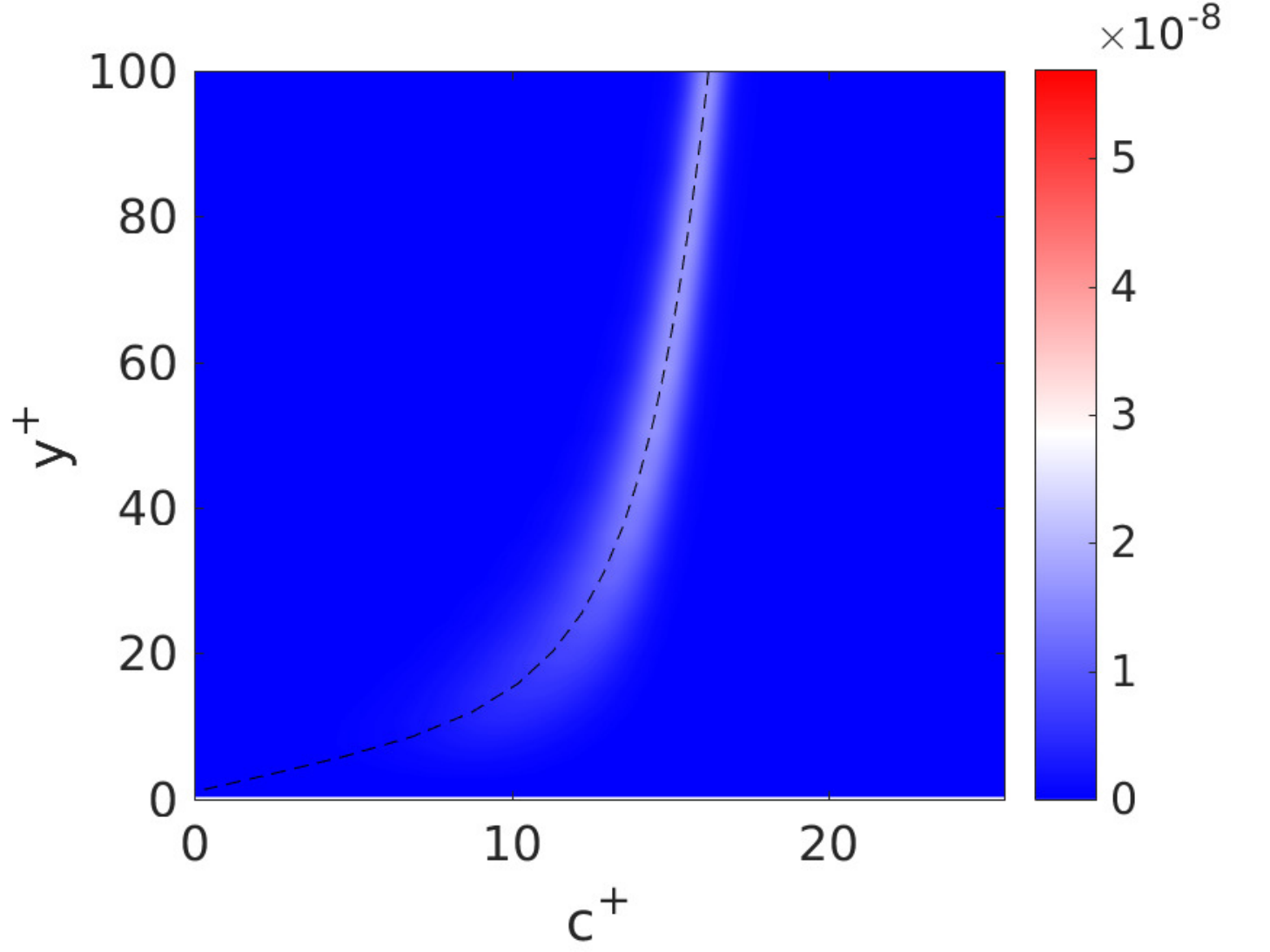} 
\put(-155,100){{$(a)$}}
 \includegraphics[width=0.39\textwidth]{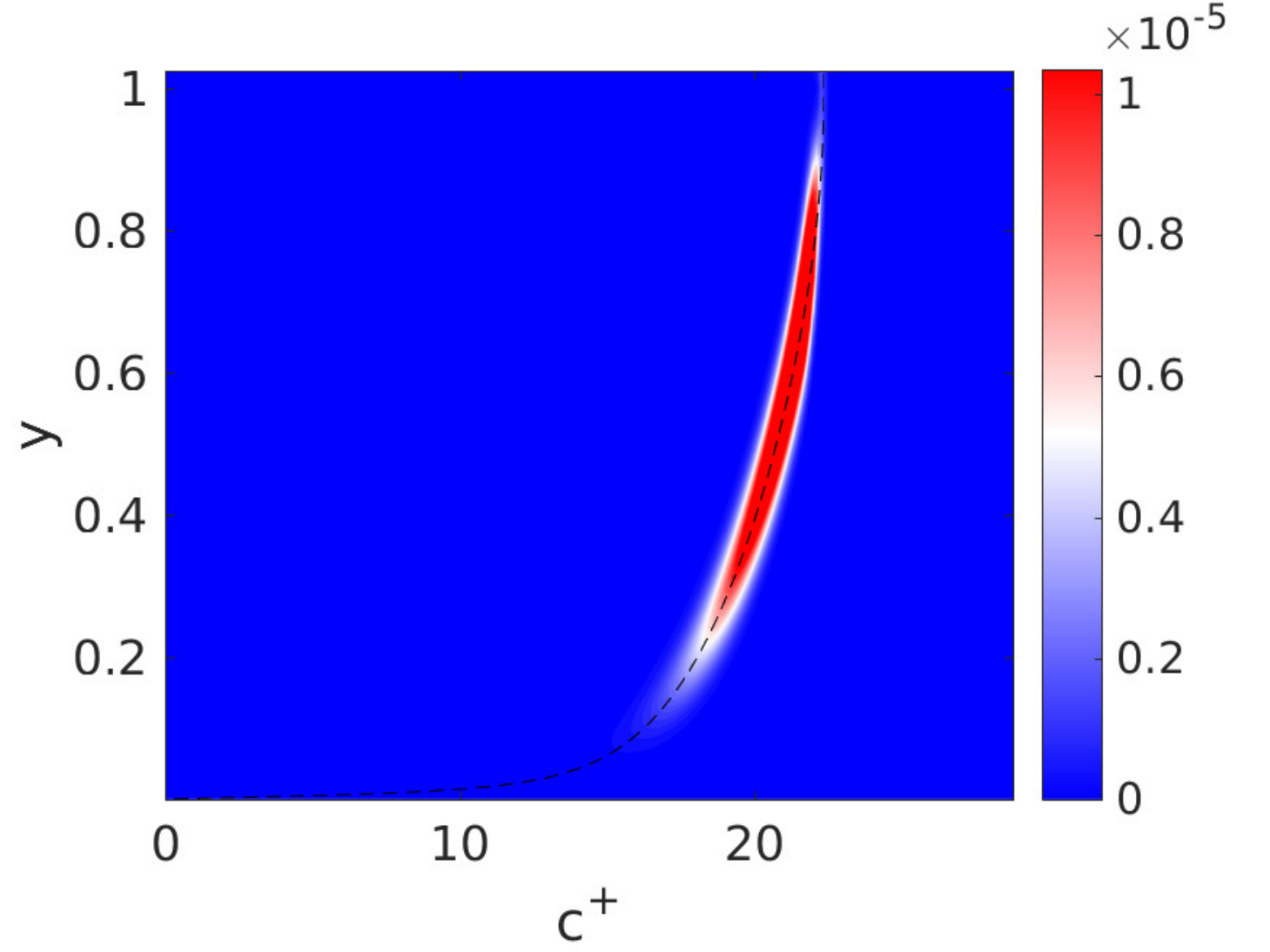} 
\put(-150,100){{$(b)$}}
}
\centerline{
\includegraphics[width=0.39\textwidth]{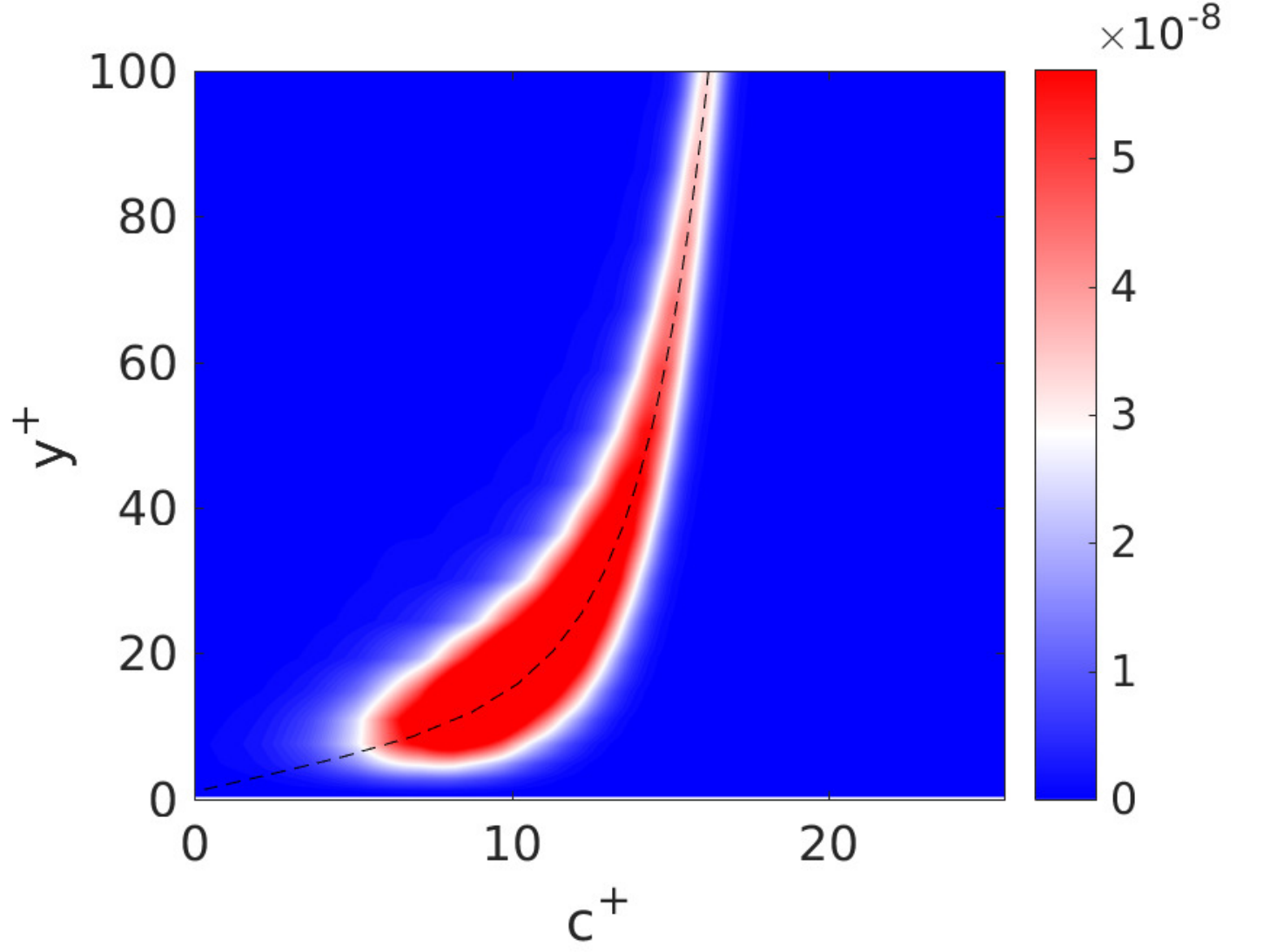} 
\put(-155,100){{$(c)$}} 
 \includegraphics[width=0.39\textwidth]{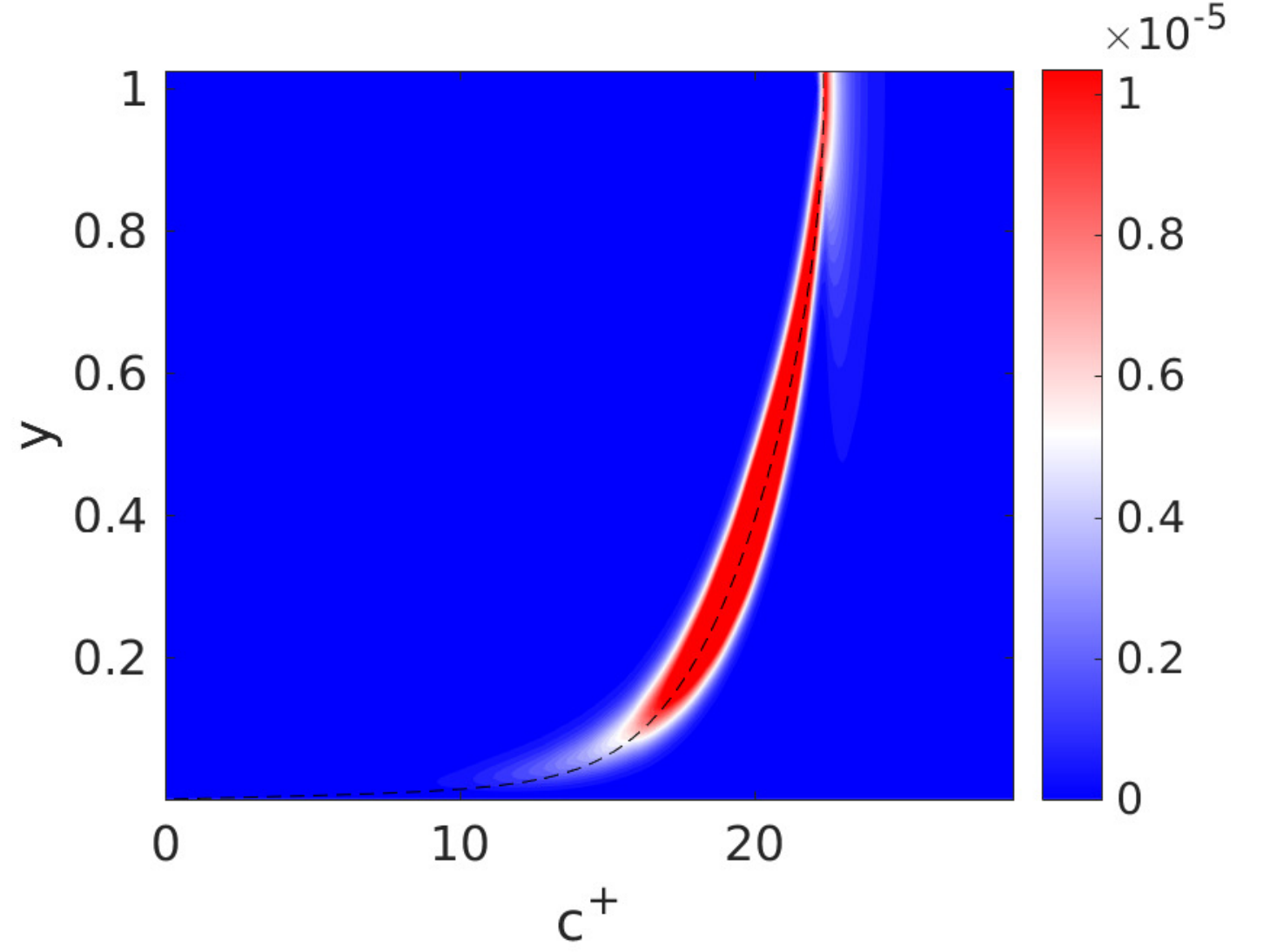} 
\put(-150,100){{$(d)$}}
}
   \vspace*{-2mm}
  \caption{Linear $\nu$ model estimation of the streamwise velocity  \spe\, $\avg{\widetilde{u}(y)\widetilde{u}^*(y)}$ versus the phase speed $c^+$ with (white-noise) flat-spectrum stochastic forcing (top panels $a$ and $b$) and coloured-noise stochastic forcing (bottom panels $b$ and $d$) for the near-wall structures with $\lambda_x^+ = 450,\lambda_z^+ = 100$ (left panels $a$ and $c$) and the large-scale structures with $\lambda_x=3, \lambda_z=1.5$ (right panels $b$ and $d$).
The mean velocity profile $U^+$ is reported as a dashed black line.
The colour-scale is the same as the one used to represent direct numerical simulations data in \reffig{Welch}.
} \label{fig:WelchlikeStoFoNu}
\end{figure}
\begin{figure}
  \centerline{
\includegraphics[width=0.39\textwidth]{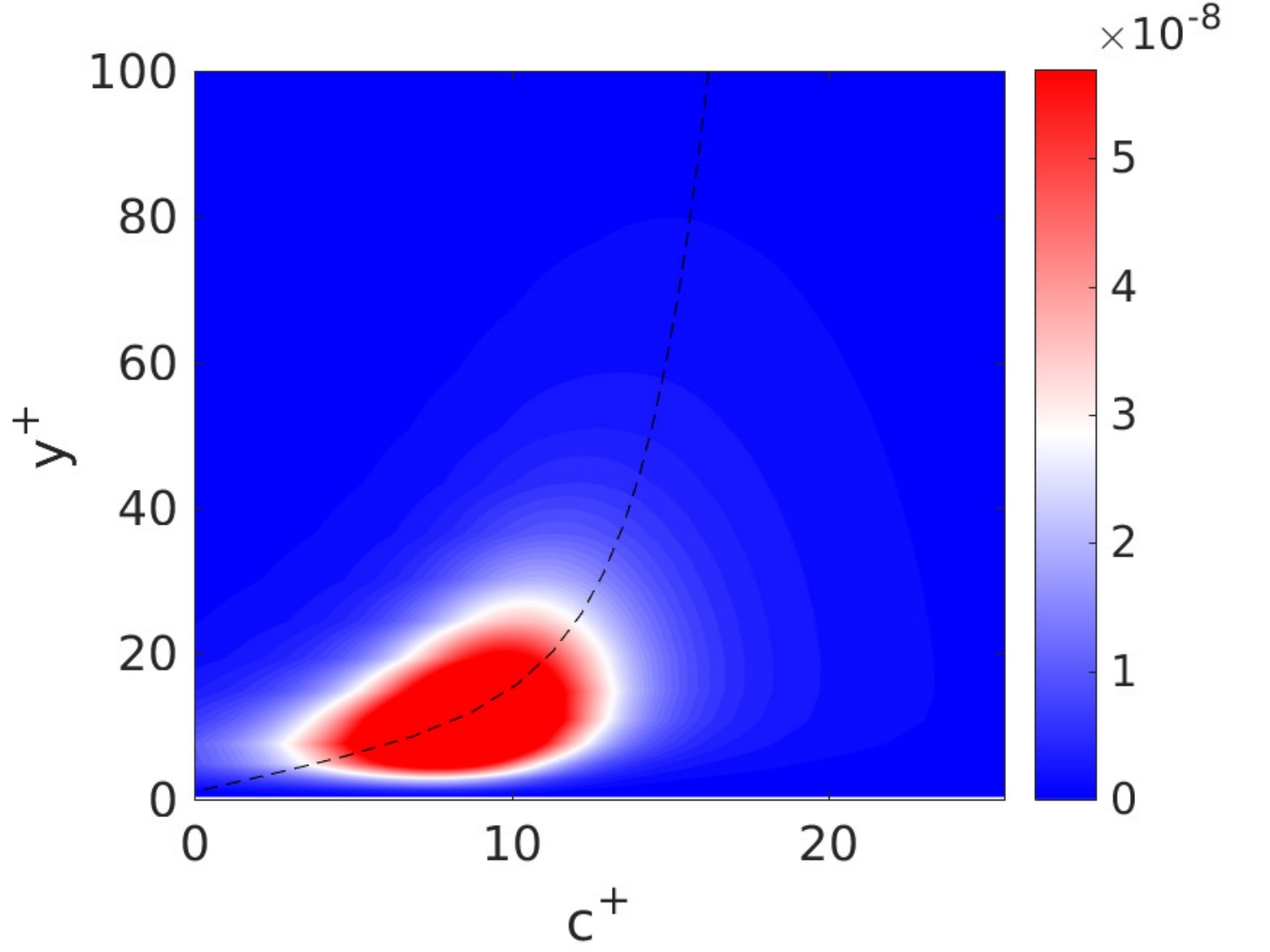} 
\put(-155,100){{$(a)$}}
 \includegraphics[width=0.39\textwidth]{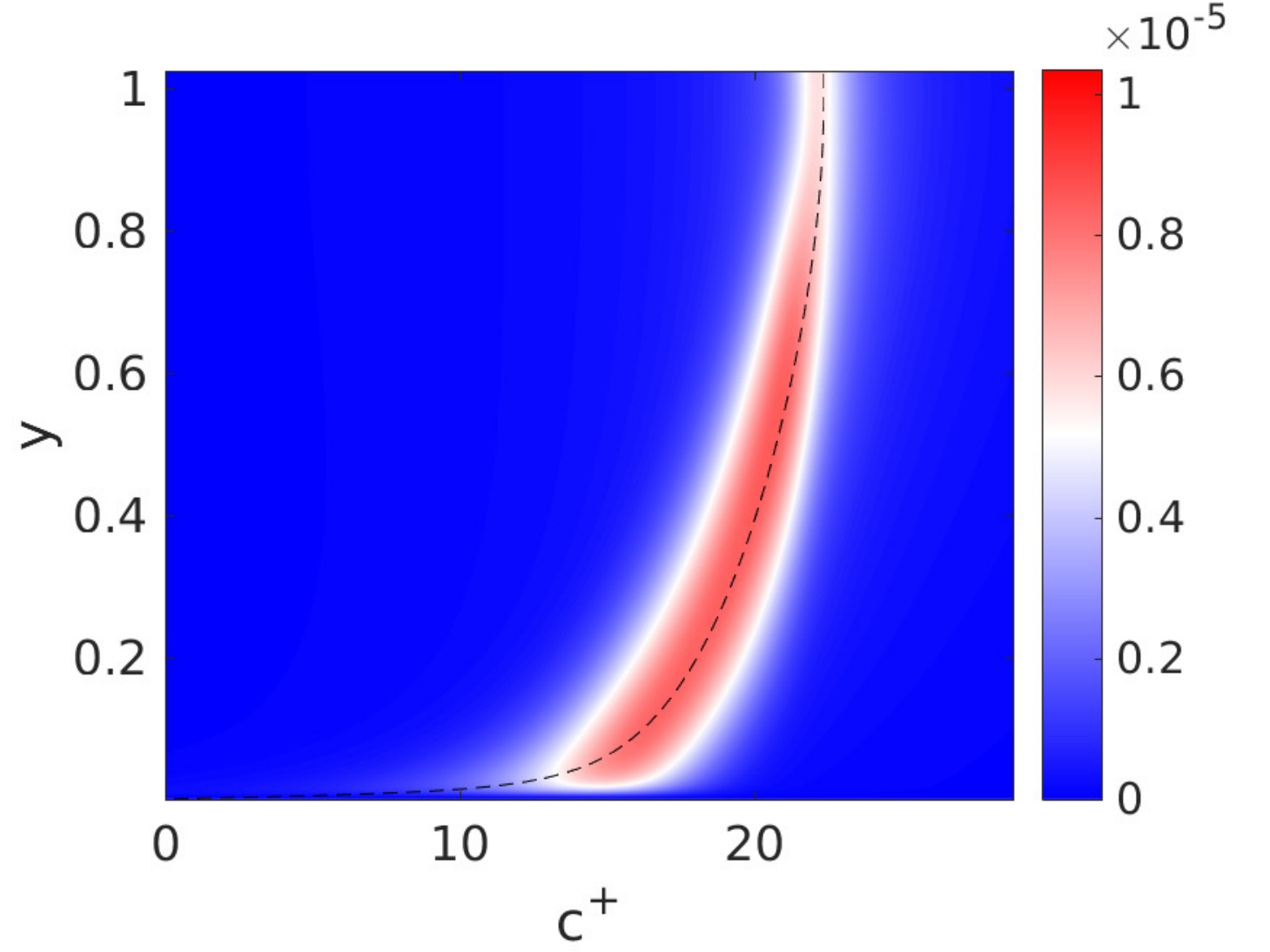}  
\put(-150,100){{$(b)$}}
}
\centerline{
\includegraphics[width=0.39\textwidth]{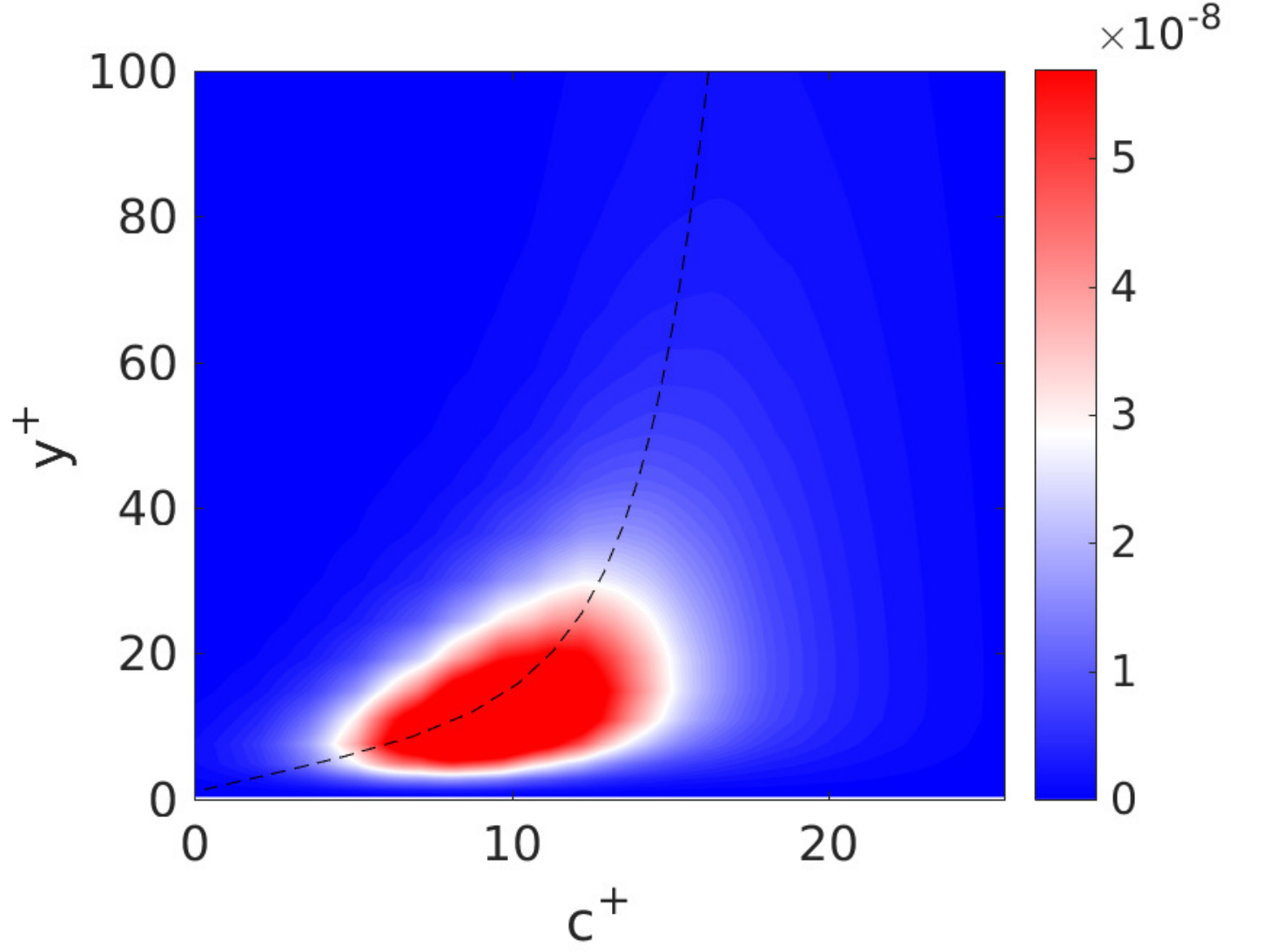}
\put(-155,100){{$(c)$}} 
 \includegraphics[width=0.39\textwidth]{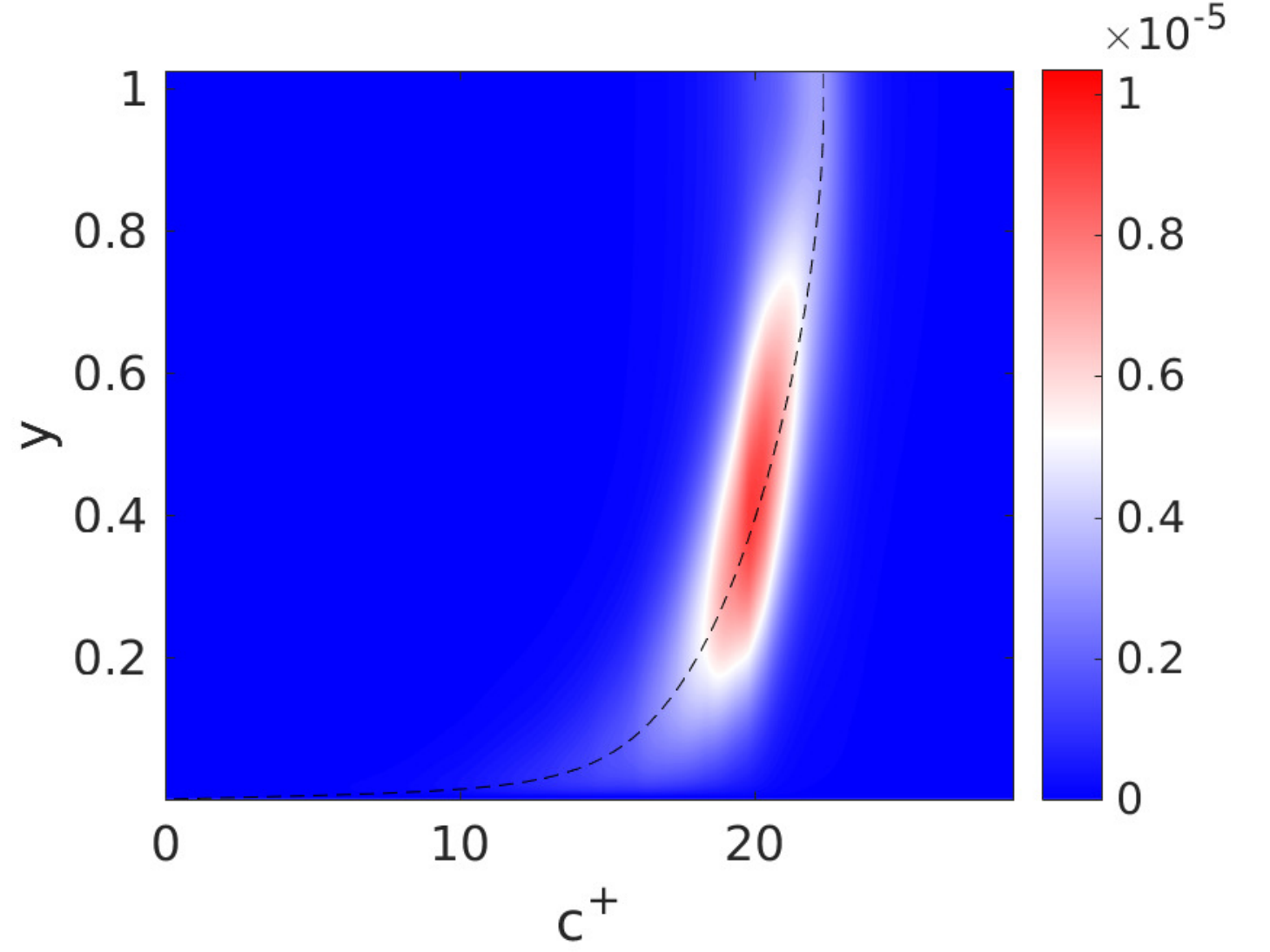} 
\put(-150,100){{$(d)$}}
}
  \vspace*{-2mm}
  \caption{Linear $\nu_t$-model estimation of the streamwise velocity  \spe\, $\avg{\widetilde{u}(y)\widetilde{u}^*(y)}$ versus the phase speed $c^+$ with (white-noise) flat-spectrum stochastic forcing (top panels $a$ and $b$) and coloured-noise stochastic forcing (bottom panels $b$ and $d$) for the near-wall structures with $\lambda_x^+ = 450,\lambda_z^+ = 100$ (left panels $a$ and $c$) and the large-scale structures with $\lambda_x=3, \lambda_z=1.5$ (right panels $b$ and $d$).
The mean velocity profile $U^+$ is reported as a dashed black line.
The colour-scale is the same as the one used to represent direct numerical simulations data in \reffig{Welch}.
} \label{fig:WelchlikeStoFoNuT}
\end{figure}

\section{Estimations of \crospe\, based on linear models}
\label{sec:LinMods}

\subsection{Estimations of the \crospe} \label{sec:FreqDep}
We now evaluate the capacity of the linear models of the coherent structures dynamics to reproduce statistics of the turbulent flow via the estimation $\Scorr^{(res)}=p \bfH \bfH^*$ given by \refeq{Sres} using both the $\nu_t$ and the $\nu$ models to build the resolvent $\bfH$. 

As a first case, we assume that the temporal power spectrum of the forcing is flat i.e. that $p$ does not depend on $\omega$ (white noise assumption) and $p=\overline{p}$ is chosen such that the estimated {\it total} spectral power of the streamwise velocity matches the one issued from direct numerical simulations
$\int_{-\infty}^ \infty \int_{-1}^1 \Scorruu^{(est)}(\alpha,y,y,\beta,\omega) dy d\omega=
\int_{-\infty}^ \infty \int_{-1}^1 \Scorruu^{(dns)}(\alpha,y,y,\beta,\omega) dy d\omega$.
As a second case, we remove the white-noise assumption and determine $p$ (which depends now on $\omega$) such that the estimated power of the streamwise velocity matches that issued from direct numerical simulations at each selected frequency $\omega$:
$\int_{-1}^1 \Scorruu^{(est)}(\alpha,y,y,\beta,\omega) dy= \int_{-1}^1 \Scorruu^{(dns)}(\alpha,y,y,\beta,\omega) dy$. We will refer to this second case  as `coloured' noise \citep[see e.g.][]{Zare2017}.

In  Figs.~\ref{fig:WelchlikeStoFoNu} and \ref{fig:WelchlikeStoFoNuT} we show,  analogously to \reffig{Welch}, the $y$-profiles of the estimated streamwise velocity \spe\, 
versus the phase speed $c^+$.

The cases where the estimation is based on the $\nu$-model to compute the resolvent are reported in \reffig{WelchlikeStoFoNu}.
For the case with white-noise (flat-spectrum) stochastic forcing (panels $a$ and $b$), the $\nu$-model does not  select the correct values and locations of the \spe\, peaks, which are predicted to be near the channel center, and therefore are predicted at too large $c^+$ values (and therefore too large $\omega$ values). 
For both near-wall and large-scale structures the \spe\, appears also to be too narrowly concentrated near the critical layer (where $U^+=c^+$) when compared to the DNS data of \reffig{Welch}.
The estimation improves for the case of coloured noise (panels $c$ and $d$), where the selective power spectrum of the forcing is able to drive the response peaks nearer their DNS values. 
However, 
even in this case the \spe\, spatial distributions remain too narrowly concentrated and therefore too large near the critical layer curves (the uniform dark red regions in panels $b$, $c$ and $d$ which are strongly offscale).

As shown in \reffig{WelchlikeStoFoNuT}, the use of the $\nu_t$ model leads to a significant improvement because the effective eddy diffusivity dampens the response near the channel centre (where the eddy viscosity is high) and smoothens the critical layer peaks.
Even with (non-selective, flat spectrum) white-noise stochastic forcing (panels $a$ and $b$), the $\nu_t$-model is able to select a reasonable $y^+_{max}$-$c^+_{max}$ location of the cross-spectral peak for buffer-layer structures.
For large-scale structures (panel $b$), however, the amplitude of the response is still over-predicted near the channel centre and near the wall.

The estimation provided by the $\nu_t$ model further improves when the coloured-spectrum forcing is used (panels $c$ and $d$).
A significant improvement is obtained, in particular, for the large-scale case (panel $d$) where the streamwise velocity power spectral density is not too large near the wall and the channel center and has a well defined peak located at reasonable $y_{max}$ and $c^+_{max}$ values.

\begin{figure}
  \centerline{
  \includegraphics[width=0.39\textwidth]{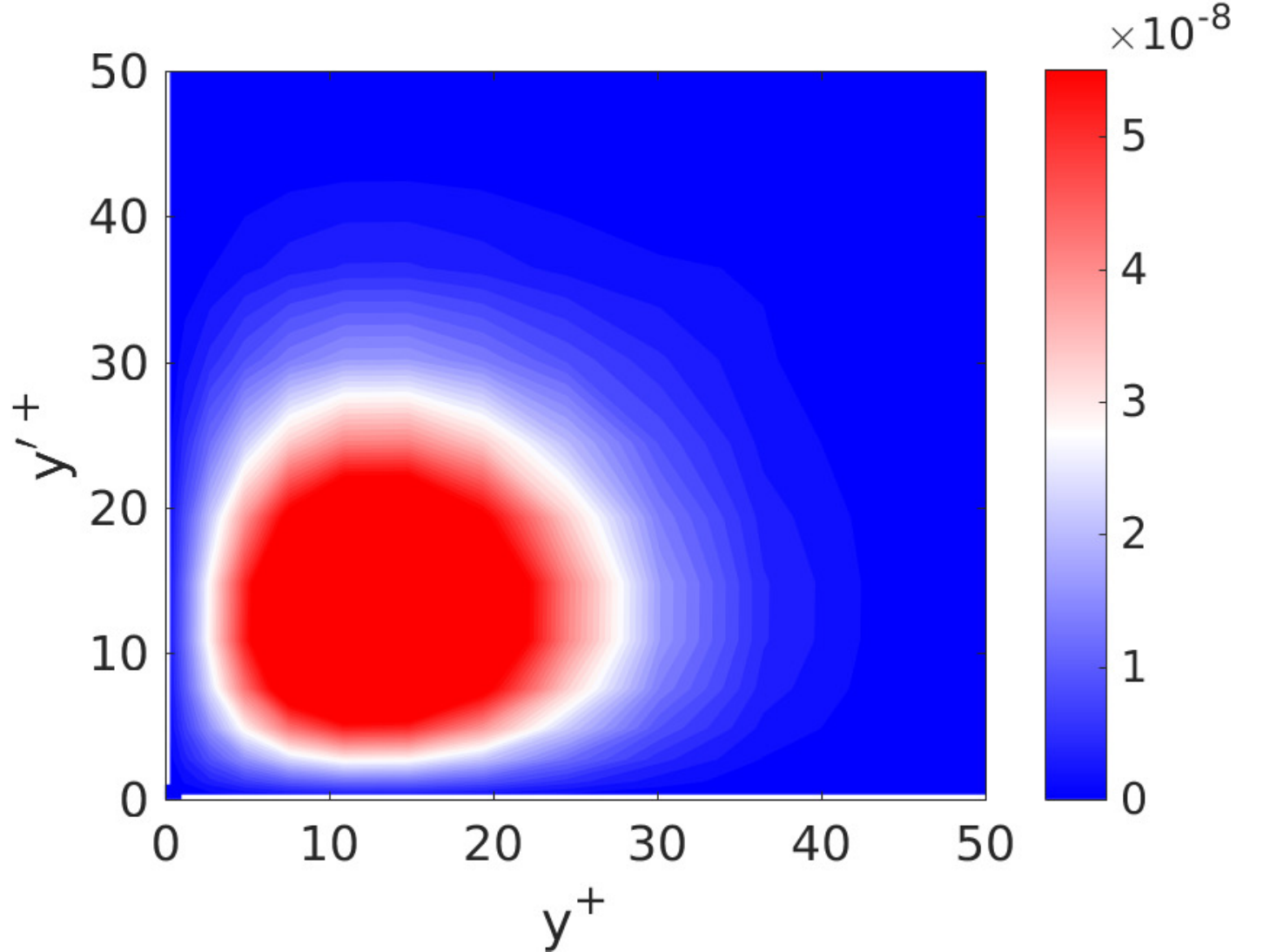} 
\put(-155,100){{$(a)$}}
  \includegraphics[width=0.39\textwidth]{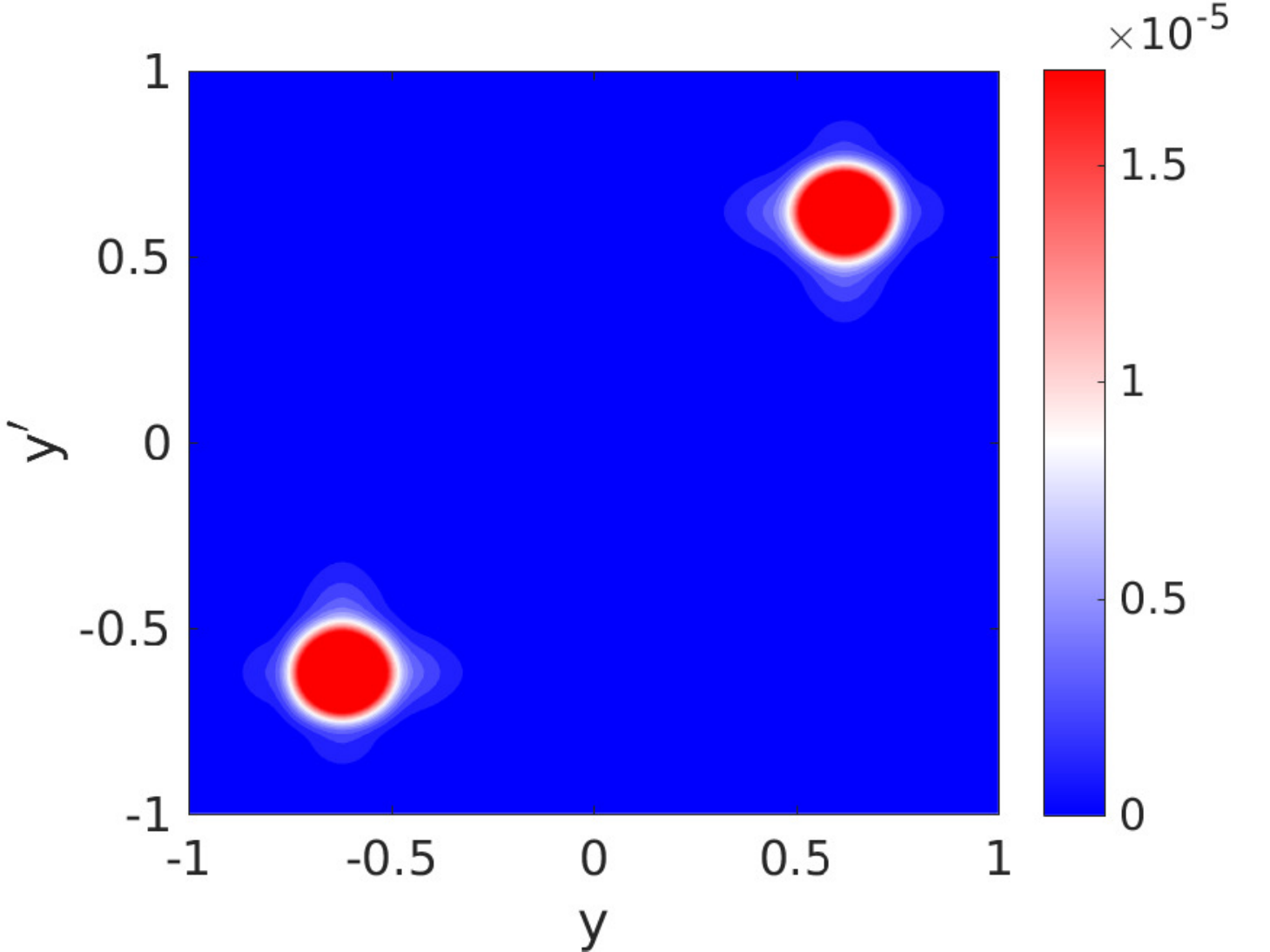}
\put(-150,100){{$(b)$}}
}
  \centerline{
  \includegraphics[width=0.39\textwidth]{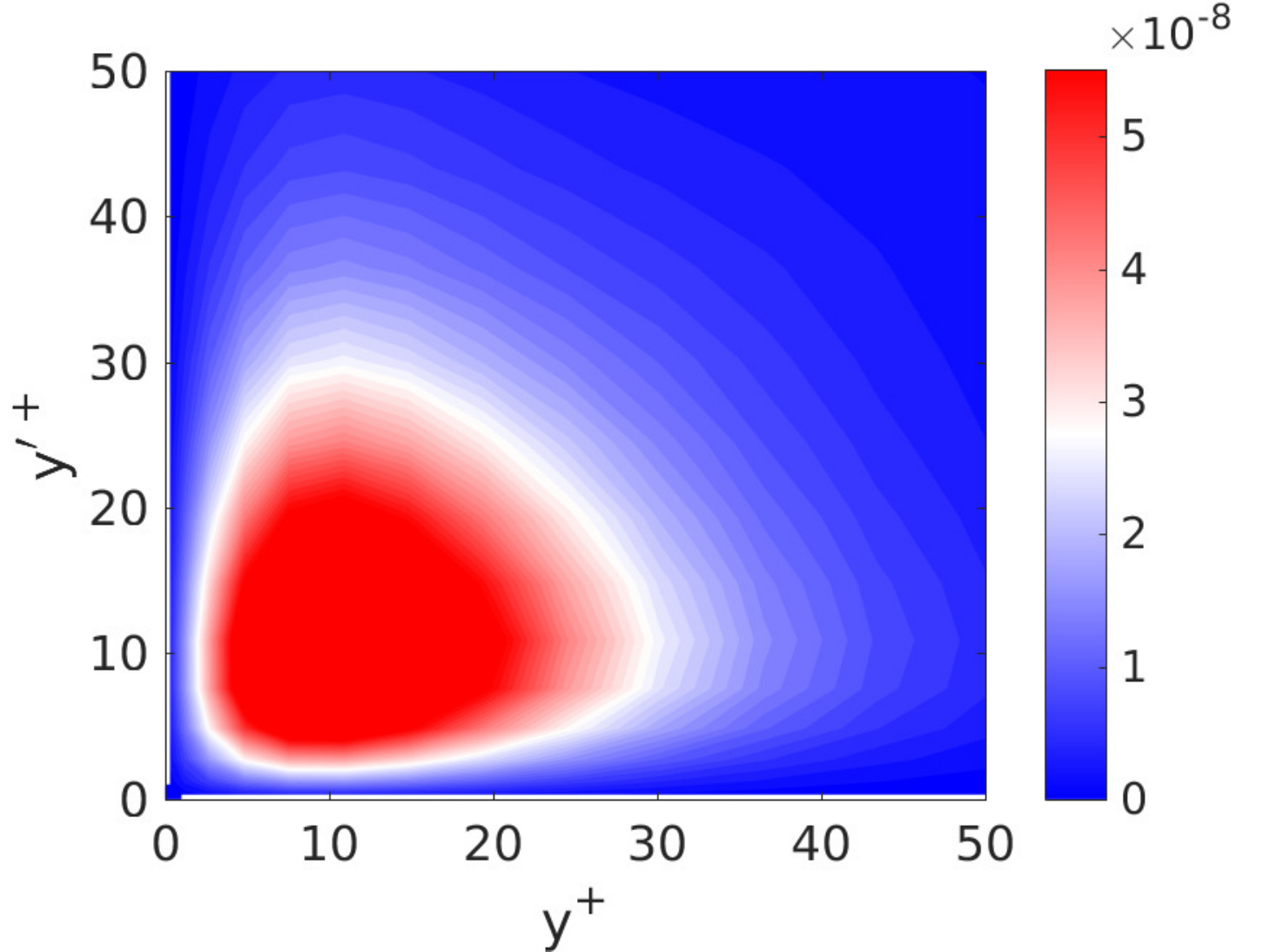} 
\put(-155,100){\small {$(c)$}}
  \includegraphics[width=0.39\textwidth]{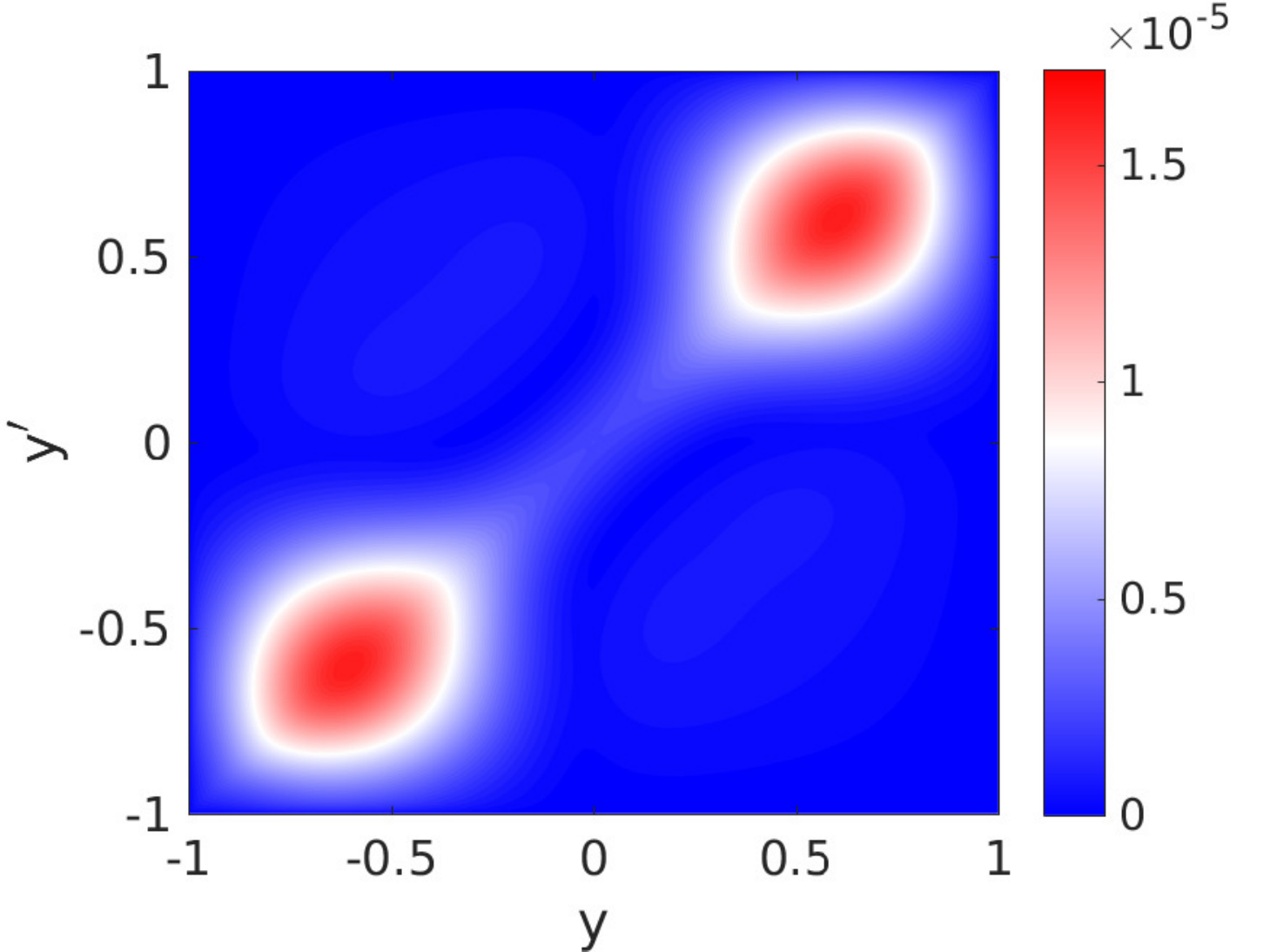}
\put(-150,100){{$(d)$}}
}
\caption{Absolute value of the streamwise velocity \crospe\, 
$|\avg{\widetilde{u}(\alpha,y,\beta,\omega_{max}) \widetilde{u}^*(\alpha,y',\beta,\omega_{max})}|=
|\Scorruu(\alpha,y,y',\beta,\omega_{max})|$ estimated by means of the $\nu$-model (top panels $a$ and $b$) and of the $\nu_t$-model (bottom panels $c$ and $d$) for 
near-wall structures with $\lambda_x^+ = 450,\lambda_z^+ = 100$ at the peak frequency $\omega=4.3$ (panels $a$ and $c$) and 
large-scale structures with $\lambda_x=3, \lambda_z=1.5$ at the peak frequency  $\omega=1.4$ (panels $b$ and $d$).
The same colour-scale as in \reffig{WelchCrossCo} is used.
} \label{fig:WelchLikeCrossCoEst}
\end{figure}
These observations are further supported by the examination of streamwise velocity \crospe\, 
$\Scorruu(\alpha,y,y',\beta,\omega)=\avg{\widetilde{u}(\alpha,y,\beta,\omega) \widetilde{u}^*(\alpha,y',\beta,\omega)}$ which are shown in \reffig{WelchLikeCrossCoEst} as a function of $y$ and $y'$ for the same $(\alpha,\beta)$ and $\omega_{max}$ values already considered in \reffig{WelchCrossCo}.
Also in this case, the $\nu$-model predicts too `narrow' and too large \crospe\, distributions which decay too fast from their peak location while the $\nu_t$ model
is able to estimate reasonably well the \crospe\, issued from the DNS data.

\subsection{Expansion on SPOD and resolvent modes} \label{sec:SPODexp}

Great attention has been given recently to the use of highly truncated expansions of the velocity \spe based on resolvent and spectral POD (SPOD) modes. 
Even if the main focus of this paper is not on these expansions, we find it useful to briefly examine the relevance of using the $\nu_t$ or the $\nu$ models in their computation.

Considering, for simplicity, only the streamwise velocity correlations, the spectral POD modes are the set of eigenfunctions $\psi_j^{(SPOD)}(y)$ of the self-adjoint operator $\Scorruu^{(dns)}$ ordered by decreasing corresponding real eigenvalues  $e_j^{(SPOD)}$  \citep{Lumley1970,Picard2000,Semeraro2016b,Towne2018}. 
These modes are orthonormal in the inner product 
associated to the power norm  
and therefore each SPOD eigenvalue, when normalized with respect to the sum of all eigenvalues, represents the fraction of the power associated with the corresponding SPOD mode.
Thus, a salient property of the SPOD modes is that they provide the orthonormal basis of functions which is optimal (in terms of spectral power) for the expansion of the streamwise velocity coherent fluctuations.

Analogously, the resolvent modes $\psi_j^{(res)}(y)$ are the set of (orthonormal) eigenfunctions of 
$\Scorruu^{(res)}$ ordered by decreasing respective real eigenvalues $e_j^{(res)}$ and, in the present contest, they represent an estimation of the SPOD modes, under the assumption that $\Pcorr=p\,\bfI$.
The leading resolvent eigenvalue $e_1^{(res)}=p \|\bfH \|^2$ also represents the optimal power amplification (modulo $p$) of the forcing supported by the resolvent operator $\bfH$ \citep[see e.g.][]{Trefethen1993,Schmid2001}.

\begin{figure}
  \centerline{
  \includegraphics[width=0.4\textwidth]{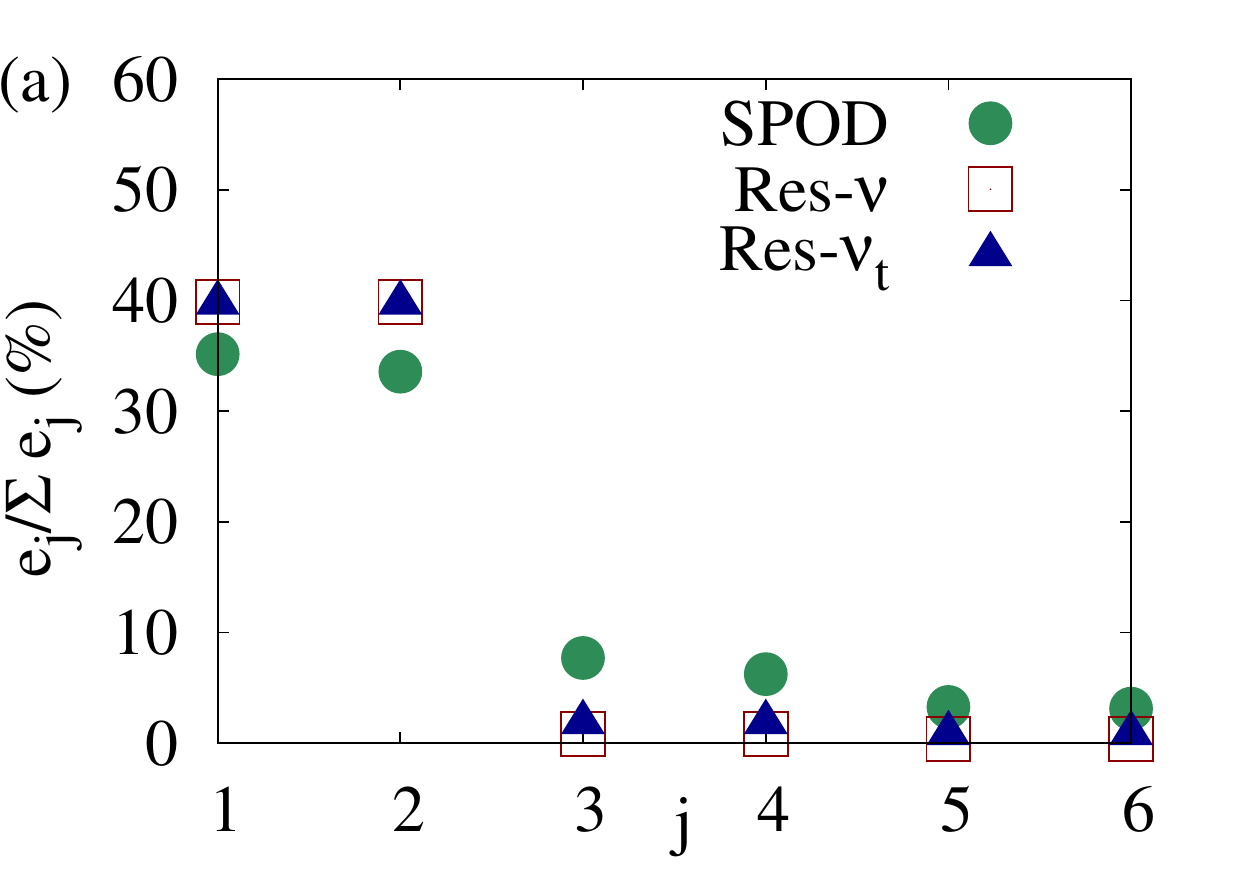} 
  \includegraphics[width=0.4\textwidth]{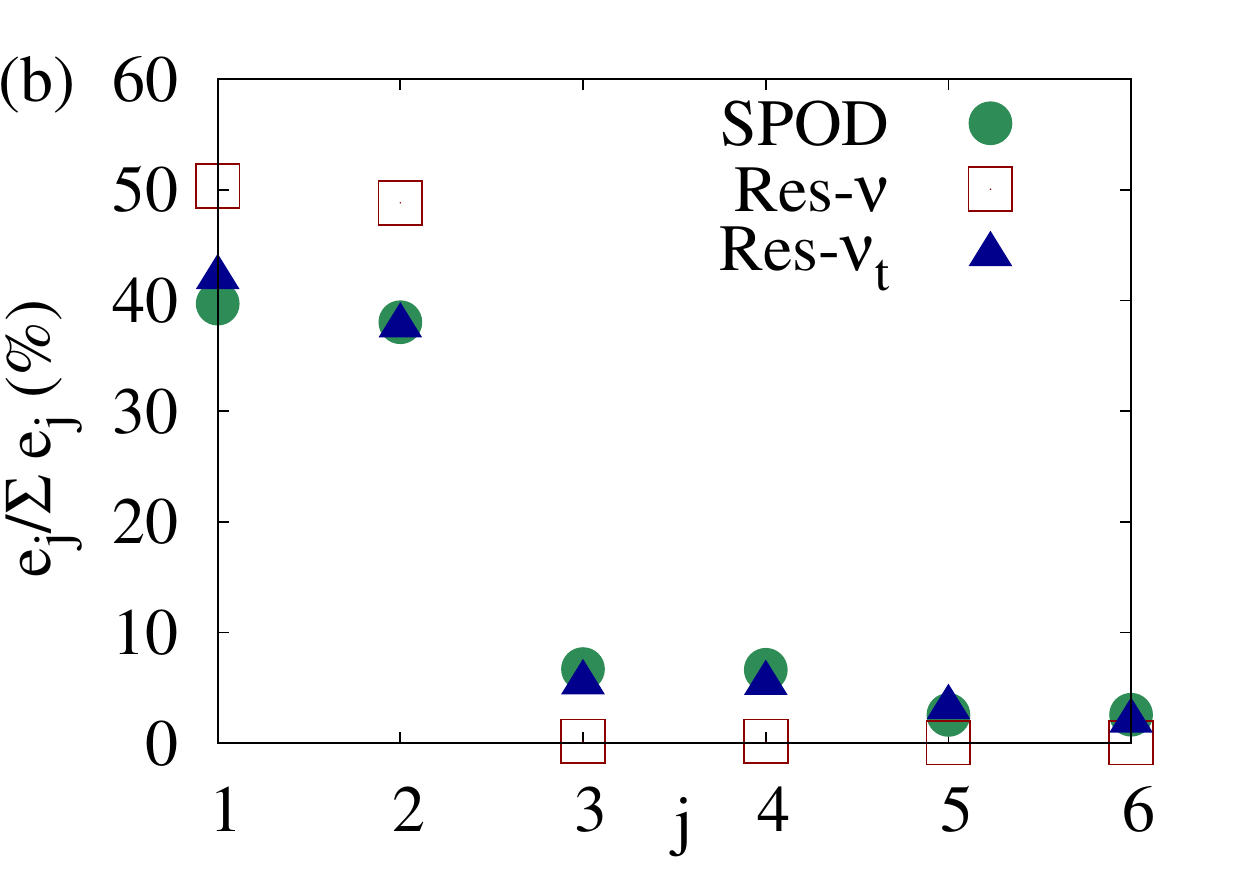} 
}
\vspace{5mm}
\caption{Streamwise velocity power content of the six leading SPOD and resolvent modes normalized by the total power for 
$(a)$~near-wall structures with $\lambda_x^+ = 450,\lambda_z^+ = 100$ at the peak frequency $\omega=4.3$ and 
$(b)$~large-scale structures with $\lambda_x=3, \lambda_z=1.5$ at the peak frequency  $\omega=1.4$.
} \label{fig:Ev_SPOD}
\end{figure}
\begin{figure}
  \centerline{
  \includegraphics[width=0.4\textwidth]{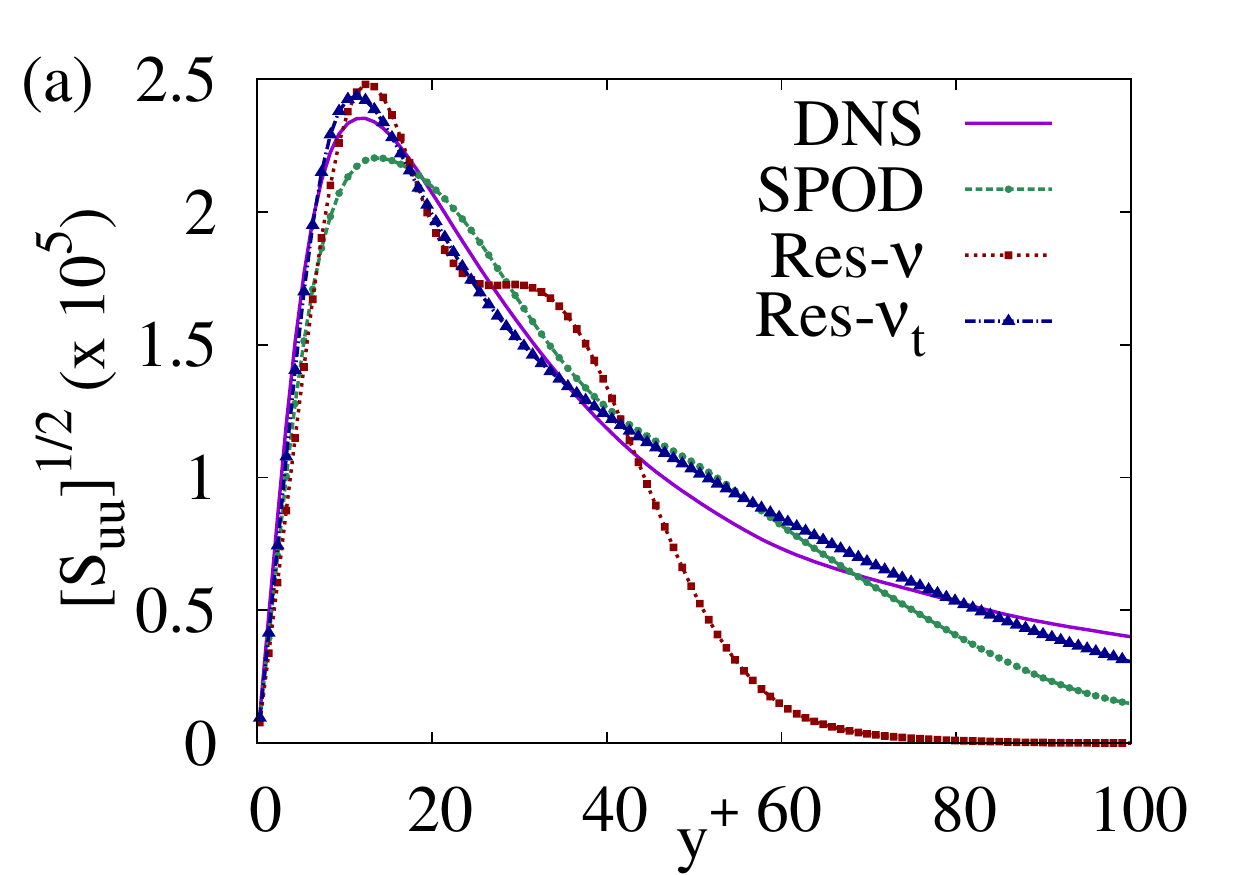} 
  \includegraphics[width=0.4\textwidth]{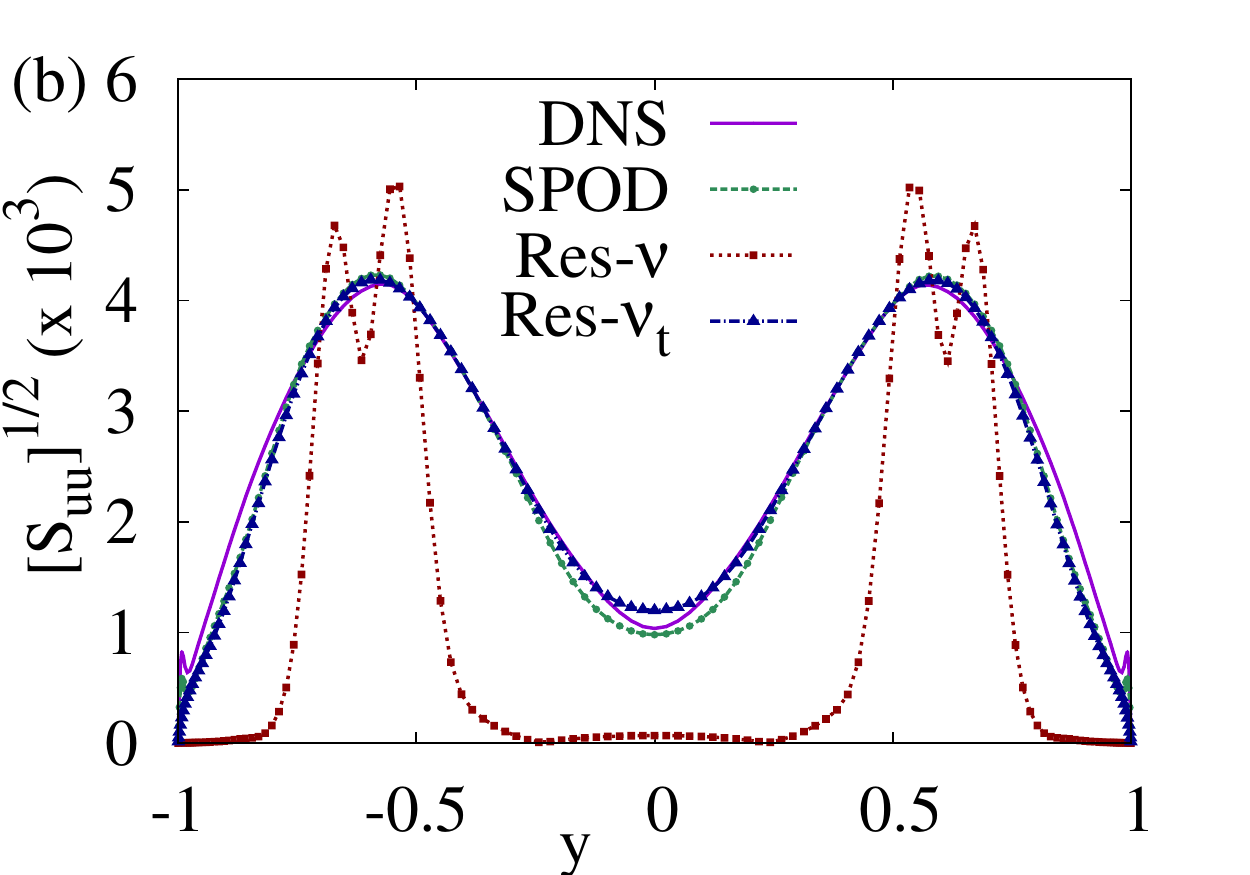} 
}
\vspace{5mm}
\caption{Comparison of the streamwise velocity root-mean-square ($rms$) profile $\widetilde{u}_{rms}^{(dns)}=[\Scorruu^{(dns)}]^{1/2}$ computed by direct numerical simulation (DNS) to six-modes expansions based on spectral proper orthogonal decomposition (SPOD) of the \spe\, computed by direct numerical simulation and six-modes expansions on resolvent modes (Res) based on the $\nu$ and the $\nu_t$ models respectively for 
$(a)$~near-wall structures with $\lambda_x^+ = 450,\lambda_z^+ = 100$ at the peak frequency $\omega=4.3$ and 
$(b)$~large-scale structures with $\lambda_x=3, \lambda_z=1.5$ at the peak frequency  $\omega=1.4$.
} \label{fig:SPOD}
\end{figure}

We therefore examine the two already considered $(\alpha,\beta)$ pairs corresponding to buffer-layer and large-scale structures and the corresponding peak frequencies $\omega_{max}$ (the same as in \reffig{WelchCrossCo} and \reffig{WelchLikeCrossCoEst}); for these values, we compute the SPOD and resolvent modes and the corresponding eigenvalues based on the $\nu$ and the $\nu_t$ models, respectively including appropriate integration weights in the discretized operators \citep[see e.g.][]{Zare2017,Schmidt2018,Towne2018}.

In what concerns the eigenvalues, a common trend is observed for the SPOD and the resolvent modes, as shown in \reffig{Ev_SPOD}: the two leading modes capture 70-90\% of the total power, depending on the model; six modes are enough to capture more than 95\% of the power of large-scale structures while they capture $\approx$85\% of the power of near-wall structures \citep[these values are similar to those found for linearly estimated space-only PODs by][]{Farrell1993b,Hwang2010,Hwang2010c}.

Finally, we expand the measured streamwise velocity root-mean-square ($rms$) profile $\widetilde{u}_{rms}^{(dns)}(y)=[\Scorruu^{(dns)}(y,y)]^{1/2}$ for the considered $\alpha,\beta,\omega_{max}$ on the leading six SPOD and resolvent modes getting the estimates
$\widetilde{u}_{rms}^{(SPOD)}(y)=\sum_{j=1}^6 a_j^{(SPOD)} \psi_j^{(SPOD)}(y)$,
$\widetilde{u}_{rms}^{(res)}(y)=\sum_{j=1}^6 a_j^{(res)} \psi_j^{(res)}(y)$.

These estimates are compared to the original profile in \reffig{SPOD};  from this figure we see that the SPOD expansion and the resolvent modes expansion based on the $\nu_t$ model provide a reasonable approximation of $\widetilde{u}_{rms}^{(dns)}$ for buffer-layer structures (\reffig{SPOD}$a$) and a quite good approximation for large-scale structures (\reffig{SPOD}$b$) while this is clearly not the case for the expansion based on the $\nu$ model resolvent modes, especially for large-scale structures.

\section{Summary and conclusions} \label{sec:concl}

In this study we have compared 
the `measured' spatio-temporal \crospe\, $\Scorr^{(dns)}$ obtained by direct numerical simulation of plane channel flow at $\Retau=1007$ to the estimated one $\Scorr^{(res)}= p \bfH \bfH^*$ based on the resolvent operator (linear transfer function) $\bfH$ under the customary assumption that the forcing spatio-temporal \crospe\, is of the type $\Pcorr=p\, \bfI$.
The main advantage of this type of analysis is that the contributions of different temporal frequencies to the spatial energy spectral density can be analysed separately, as emphasized in the recent studies of \cite{Semeraro2016b,Schmidt2018} and \cite{Towne2018} who analyse the relation between proper orthogonal decomposition and resolvent analysis.

Two cases have been considered for the forcing power spectrum $p(\omega)$: a flat (white-noise) forcing power spectrum $p=\overline{p}$ and that of a (`coloured' noise) forcing power spectrum where $p(\omega)$ is chosen so as to match the power spectrum of 
the estimated streamwise velocity \spe\, 
 to the measured one.
Two separate linear models have been used to evaluate the resolvent: the first, labelled `$\nu_t$ model', includes the effect of the turbulent Reynolds stresses modelled with an eddy viscosity while the second, labelled `$\nu$ model', does not include them.

The two linear models and the two types of forcing power spectrum, have been tested for structures with spatial scales typical of buffer-layer structures ($\lambda_z^+=100, \lambda_x^+=450$) and of large-scale motions ($\lambda_z= 1.5, \lambda_x= 3$). 
We find that estimations based the $\nu_t$ model resolvent lead to  qualitatively correct estimations of \crospe\, distributions and of the dependence of the \spe\, wall-normal distributions on the frequency (or, equivalently, the phase speed).
On the contrary, estimations based on the $\nu$ model lead to generally overestimated values of the streamwise velocity \spe\,  which are (too) narrowly concentrated near the critical layer. 
The $\nu$ model is also unable to even qualitatively reproduce the 
$c^+_{max}-y^+_{max}$ 
values of the peak
amplitude of the \spe\, for buffer-layer structures. 
We also show that in all cases, as expected, the estimation is improved by using the appropriately coloured input power spectrum $p(\omega)$ instead of white noise as already observed by \cite{Zare2017} using a different methodology. 

Similar results are found when comparing highly truncated expansions of the measured streamwise velocity \spe\, on the basis of the leading spectral POD (SPOD) modes to the expansions on leading resolvent modes.
The six-modes expansion based on SPOD modes approximates well the original \spe\, and a comparable accuracy is obtained by making use of the same number of resolvent modes based on the $\nu_t$ model, while the use of resolvent modes based on the $\nu$ model leads to a less accurate approximations, especially for large-scale structures.

These results can be understood by recalling that in estimates based on the $\nu_t$ model the effect of turbulent Reynolds stresses is included in the resolvent $\bfH$ while in the $\nu$ model the effect of turbulent Reynolds stresses resides only in the forcing and the associated \crospe\, $\Pcorr$. 
As a consequence, when using linear models in the estimation $\Scorr^{(est)}=\bfH \Pcorr \bfH^*$, when no accurate information on $\bfP$ is available (which is often the case, especially at high Reynolds numbers, and one therefore assumes the standard $\Pcorr=p\bfI$), the $\nu_t$ model performs relatively well because the effect of the turbulent Reynolds stresses is embedded in $\bfH$. 
It is likely that the $\nu$ model could perform equally well in the case where relevant information on turbulent Reynolds stresses is available leading to a realistic modelling of the forcing cross-spectral tensor $\Pcorr$;
in this second case the scale selection would be dictated by  $\Pcorr$ instead of $\bfH$.

These findings are consistent with those of \cite{Illingworth2018} who find that the use of the $\nu_t$ model leads to a better performance in the estimation of streamwise velocity fluctuations in the turbulent channel at the same friction Reynolds number $\Retau=1007$ considered here (however, they assume a white-noise spectrum and do not separate the different frequency contributions in their analysis).
It is also important to note that there is no contradiction between our findings and those of \cite{Semeraro2016b,Schmidt2018} and \cite{Towne2018} who show a reasonable agreement of resolvent modes computed with the $\nu$ model to experimental \crospe\, data in turbulent jets because, contrary to turbulent wall-bounded flows, in the case of jets the turbulent eddy viscosity has only a weak dependence on the (cross-stream) radial coordinate and therefore the $\nu$ and $\nu_t$ formulations almost coincide except for a Reynolds number rescaling by $\nu/\nu_T$.

In the present investigation turbulent Reynolds stresses have been modelled with eddy viscosity, as in a number of previous studies, but an improvement of quantitative predictions of the \crospe\, tensor could probably come from a more advanced modelling of the Reynolds stress tensor in the linear operator.
Also, as the linear operators that we have discussed do mainly model the non-normal streaks amplification mechanism,
important progress could come from an improved, necessarily nonlinear, modelling of the {\it coherent} forcing terms which are related to the regeneration mechanism of the vortices in the self-sustained processes discussed by \cite{Waleffe1995,Hwang2010b,Hwang2011} and \cite{Cossu2017} among others.
These issues are under current intensive scrutiny.

\appendix
\section{Linear model operators}
\label{app:ABCD}

The linear system of \refeq{DynSys} is derived \citep[see e.g.][]{Schmid2001} from  equations~(\ref{eq:LinNS})  with the operators $\bfA$ and $\bfB$ defined as
\begin{eqnarray}
\label{eq:OSSQ}
\bfA=\left[\begin{array}{cc}
   \Delta^{-1}\mathcal{L_{OS}} & 0 \\
   -i\beta U' & \mathcal{L_{SQ}}  \\
\end{array}\right],~~~
\bfB = \left[\begin{array}{ccc}
   -i\alpha \Delta^{-1}\calD & - k^2 \Delta^{-1} & -i\beta\Delta^{-1} \calD \\
   i\beta & 0 & -i\alpha 
\end{array}\right]
\end{eqnarray}
where the generalised Orr-Sommerfeld and Squire operators \cite[]{Cossu2009,Pujals2009} are:
\begin{eqnarray}
\label{eq:OSSQop}
\mathcal{L_{OS}}&=&-i\alpha(U \Delta-U'')
       +\nu_T \Delta^2+2 \nu_T' \Delta \calD +\nu_T''(\calD^2+k^2),\\
\mathcal{L_{SQ}}&=&-i\alpha U + \nu_T \Delta+\nu_T' \calD,
\end{eqnarray}
with $\calD$ and $'$ denoting $d/d y$,  $k^2=\alpha^2+\beta^2$ and $\Delta=\calD^2-k^2$.
Homogeneous boundary conditions are enforced on both walls: $ \hat{v}(\pm 1)=\calD\hat{v}(\pm h)=\hat{\omega}_y(\pm 1)=0$.
The linear operators relating $\velhat$ and $\statehat$ are
\begin{eqnarray}
\bfC = {1\over k^2} \left[\begin{array}{ccc}
   i\alpha \calD &  -i\beta \\
   k^2 &        0      \\
    i\beta\calD & i\alpha \\
\end{array}\right]
~~~~~~;~~~~~~~
\bfD=
\left[\begin{array}{ccc}
   0 		& 1 & 0 	\\
   i\beta 	& 0 & -i\alpha	\\
\end{array}\right].
\end{eqnarray}

\section{Direct numerical simulations and data analysis}
\label{app:DNSmeth}

The simulations have been performed using the \texttt{SIMSON} code, which is pseudo-spectral and uses Fourier expansions in the streamwise and spanwise directions, where periodicity is enforced, and  Chebyshev expansions in the wall-normal direction to solve the three-dimensional time-dependent incompressible Navier-Stokes equations in the channel \citep[see][ for further details]{Chevalier2007}.

$N_x=128, N_y=129, N_z = 128$ points (before 3/2-rule dealiasing of Fourier modes) have been used for the $\Retau=1007$ simulations in the domain $L_x=3, L_y=2, L_z=1.5$, with a uniformly spaced grid in
$x$ and $z$ and Gauss-Lobatto points in $y$. At $\Retau=1007$ this corresponds to a grid spacing $\Delta x^+ = 23, \Delta z^+= 12$ and to $\Delta y^+_{min}= 0.3$ near the wall and $\Delta y^+_{max}=24$ at the channel centre.
This chosen grid is coarser than those typically used in DNS at this Reynolds number \citep[e.g.][who use $\Delta x^+ = 11, \Delta z^+= 5$]{Lee2015} to keep the data analysis manageable but it has been verified that this has no significant impact on the first ans second-order flow statistics (see \reffig{MeanFlow}).

The initial transient of the simulation is discarded and statistics and samples are accumulated starting from $t = 95000$. 
Welch's method with  Hamming windowing and $75$\% overlap has been used to compute the velocity \crospe\,  $\Scorr$ using a total of $N_s=8001$ snapshots of the DNS solutions with sampling interval  $(\Delta t)_{s}=0.25$ for a total acquisition time $T_{max}= 2000$ for the small scales, and using $N_s=20001$ snapshots sampled every $(\Delta t)_{s}=0.5$ for a total acquisition time $T_{max}= 10000$ for the large scales.  
The average temporal step of the DNS during the acquisition time is $\overline{\Delta t} =0.0106$, i.e. $\overline{\Delta t}^{\,+} =0.35$.
Data have also been averaged between the two walls.

\begin{acknowledgements}

Direct numerical simulations were performed on resources provided by the Swedish National Infrastructure for Computing (SNIC) at NSC, HPC2N and PDC
\end{acknowledgements}


\newcommand{\noopsort}[1]{} \newcommand{\printfirst}[2]{#1}
  \newcommand{\singleletter}[1]{#1} \newcommand{\switchargs}[2]{#2#1}

\end{document}